\documentclass[aps,superscriptaddress]{revtex4}
\usepackage{multirow} 
\usepackage{amssymb,amsmath,epsfig}
\usepackage{subfigure}
\usepackage{graphicx}
\usepackage{latexsym}
\usepackage{amsfonts}
\usepackage{booktabs}
\usepackage{color}
\usepackage[colorlinks=true, pdfstartview=FitV, linkcolor=blue, citecolor=red, urlcolor=magenta, breaklinks=true]{hyperref}
\usepackage{orcidlink}

\begin{document}
\title{Quasinormal modes and shadow of Schwarzschild-Tangherlini black hole with GUP}

\author{J. A. V. Campos \orcidlink{0000-0002-4252-2451}}\email{joseandrecampos@gmail.com}
\affiliation{Departamento de F\'{\i}sica, Universidade Federal de Campina Grande
Caixa Postal 10071, 58429-900 Campina Grande, Para\'{\i}ba, Brazil}

\author{M. A. Anacleto \orcidlink{0000-0003-4625-7322}}
\email{anacleto@df.ufcg.edu.br}
\affiliation{Departamento de F\'{\i}sica, Universidade Federal de Campina Grande
Caixa Postal 10071, 58429-900 Campina Grande, Para\'{\i}ba, Brazil}
\affiliation{Unidade Acad\^emica de Matem\'atica, Universidade Federal de Campina Grande
\\
58429-900 Campina Grande, Para\'{\i}ba, Brazil}

\author{F. A. Brito \orcidlink{0000-0001-9465-6868}}\email{fabrito@df.ufcg.edu.br}
\affiliation{Departamento de F\'{\i}sica, Universidade Federal de Campina Grande
Caixa Postal 10071, 58429-900 Campina Grande, Para\'{\i}ba, Brazil}
\affiliation{Unidade Acad\^emica de Matem\'atica, Universidade Federal de Campina Grande
\\
58429-900 Campina Grande, Para\'{\i}ba, Brazil}

\author{E. Passos \orcidlink{0000-0003-1718-6385}}\email{passos@df.ufcg.edu.br}
\affiliation{Departamento de F\'{\i}sica, Universidade Federal de Campina Grande
Caixa Postal 10071, 58429-900 Campina Grande, Para\'{\i}ba, Brazil}
\affiliation{Unidade Acad\^emica de Matem\'atica, Universidade Federal de Campina Grande
\\
58429-900 Campina Grande, Para\'{\i}ba, Brazil}

\author{Amilcar R. Queiroz \orcidlink{ 0000-0002-4785-5589}}\email{amilcarq@df.ufcg.edu.br}
\affiliation{Departamento de F\'{\i}sica, Universidade Federal de Campina Grande
Caixa Postal 10071, 58429-900 Campina Grande, Para\'{\i}ba, Brazil}

\begin{abstract}  

In this work, we investigate quasinormal modes and the shadow of a Schwarzschild-Tangherlini black hole in $(d+1)$ dimensions, incorporating corrections arising from the Generalized Uncertainty Principle (GUP). Using the WKB approximation up to sixth order, we calculate the quasinormal frequencies for $l=1, 2$ and $d=3, 4, 5, 6$, and analyze the effects of the dimensionless linear and quadratic GUP parameters $\bar{\alpha}$ and $\bar{\beta}$, respectively. We find that the linear correction increases the oscillation frequency and the damping rate, whereas the quadratic correction produces the opposite behavior within the parameter range considered.
Using the finite difference scheme, we obtain the time-domain evolution and find that higher dimensions lead to faster damping and that the visibility of GUP-induced modifications to the time-domain waveform depends on the excitation profile of the initial perturbation, whereas the underlying quasinormal spectrum remains determined by the black hole geometry.
Furthermore, we investigate the shadow and establish its correspondence with the quasinormal spectrum in the eikonal limit through the properties of the unstable circular null geodesics. Finally, we confront the predicted shadow size with Event Horizon Telescope observations of Sagittarius A* and use these constraints to investigate the allowed GUP parameter space.

\end{abstract}

\maketitle
\pretolerance10000

\section{Introduction}
Black holes offer a natural setting for investigating gravity in the strong field regime, allowing properties of spacetime geometry to be probed through observables associated with matter dynamics, electromagnetic radiation, and field perturbations. 
An important motivation in the experimental field, worth highlighting, is the first detection of gravitational waves by the LIGO-VIRGO collaboration~\cite{abbott2016tests} presenting results of the detection of gravitational waves from the merger of a binary black hole system. Currently, the KAGRA detector (\textit{Kamioka Gravitational Wave Detector}) is in operation, expanding the ability to observe and characterize signals from compact systems~\cite{abac2025gw250114}. Recently, the GW250114 event provided a significant demonstration of the potential of black hole spectroscopy~\cite{abac2026black}. In recent years, the results obtained by the EHT \textit{Event Horizon Telescope} have revealed, through observations, the shadow of the massive black hole located at the center of the galaxy M87~\cite{akiyama2019first, eventhorizon2019first} and, subsequently, at the center of our own galaxy, Sagittarius A*~\cite{akiyama2022first}. 
The observation of gravitational waves has opened a new window to investigate the dynamics of compact objects and test general relativity beyond the weak field regime. In particular, it is expected that the post-merger relaxation phase of these compact objects will be dominated by a superposition of damped quasinormal modes, whose complex frequencies are determined by the properties of spacetimes. Measurements of these frequencies are a direct means of testing deviations from the predictions of general relativity.

The physics of black holes in spacetime with extra dimensions has aroused increasing interest in recent years. One such solution is described by Schwarzschild–Tangherlini geometry, which generalizes the Schwarzschild solution to the dimensions $D=d+1$ of spacetime~\cite{Tangherlini:1963bw}. Dimensionality modifies the horizon scale, the effective potential for perturbations, the stability properties of null geodesics, and consequently the spectrum of quasinormal modes. Thus, in higher-dimensional black holes, Hawking radiation was investigated in~\cite{Feng:2015jlj}, while scattering and absorption properties were studied in~\cite{Tsukamoto:2014dta, Anacleto:2023ntm}, as well as quasinormal modes in~\cite{Cardoso:2003vt, Konoplya:2011qq, Sadhu:2018zyh, Matyjasek:2021xfg, Benda:2025tni}. In particular, Schwarzschild–Tangherlini black holes exhibit characteristic changes in both oscillation frequencies and damping rates of their perturbations. In this context, we can investigate how effects arising from short-distance corrections through a generalized uncertainty principle can alter the perturbation spectrum. 

Although we do not yet have a complete theory of quantum gravity, several approaches suggest that the structure of spacetime may undergo non-trivial corrections at short distances. One way to incorporate such effects is through the generalized uncertainty principle (GUP), in which the standard Heisenberg uncertainty relation is modified by terms that become relevant at high energies. Several theoretical structures of quantum gravity have been investigated~\cite{Kempf:1994su, Ali:2009zq}.
GUP corrections have been employed to investigate quantum modifications to black hole geometries~\cite{Anacleto:2020lel, Al-Badawi:2024cby, Barman:2024hwd}. These corrections also lead to specific modifications in both the quasinormal modes and the shadow of Schwarzschild black holes~\cite{Anacleto:2021qoe,Karmakar:2022idu,Chen:2023wkq,Lambiase:2023hng}. Furthermore, GUP effects on absorption and scattering have also been investigated for Schwarzschild-Tangherlini black holes~\cite{Anacleto:2023ntm}. It is worth mentioning that connections between quasinormal modes and black hole shadow have also been investigated in other quantum-corrected geometries~\cite{Liu:2020ola,Campos:2021sff, Konoplya:2024lch,Pedrotti:2024znu}.

Another important phenomenon is the shadow of a black hole, which provides a complementary analysis of the same strong field geometry. In a static, spherically symmetric spacetime, the shadow is determined by unstable circular null geodesics, directly linked to the critical impact parameter of photon trajectories. This connection becomes important when combined with the eikonal limit of the quasinormal spectrum, in which the real and imaginary parts of the quasinormal frequencies are governed, respectively, by the angular frequency and the Lyapunov exponent of the unstable circular null orbit~\cite{Cardoso:2008bp}. Consequently, the shadow and the quasinormal spectrum encode information about the geometry near the photon sphere. 
The shadow of the Schwarzschild-Tangherlini black holes has been investigated in~\cite{Singh:2017vfr}, while recent studies have demonstrated that horizon scale observations can also be used to probe potential signatures of extra dimensions in black hole shadows~\cite{Lemos:2024wwi}.
In particular, Vagnozzi et al.~\cite{Vagnozzi:2022moj} showed that the shadow size can be used to constrain a broad class of alternative geometries by comparing the theoretically predicted shadow with the observationally inferred shadow size. More recently, rotating black holes with GUP corrections were investigated by shadow measurements~\cite{Gulia:2025tvx}, and analyses were also performed for other quantum-corrected rotating geometries using observations of M87 and Sgr A*~\cite{Ali:2024ssf,Raza:2025ohk,Ahmed:2025boj}.
This provides motivation for applying the same parametric strategy to higher dimensional black holes with GUP corrections.

Many studies on GUP-induced modifications to black hole quasinormal modes and shadows have focused on four-dimensional or rotating black hole geometries.
In this work, we investigate the quasinormal modes and the time-domain response of a Schwarzschild–Tangherlini black hole with GUP correction in dimensions ($D=d+1$). We consider the linear and quadratic parameters of the GUP and analyze their effects for different dimensions of spacetime. Quasinormal frequencies are calculated using the sixth-order WKB approximation, while the numerical evolution of scalar perturbations is obtained in the time-domain using the finite difference scheme~\cite{Gundlach:1993tn}. This combination allows us to distinguish modifications in the intrinsic quasinormal spectrum from alterations in the observable waveform associated with the excitation of the perturbation.

In addition, we investigate how the spacetime dimensions and GUP parameters affect the black hole shadow, deriving the radius of the photon-unstable circular orbit and the corresponding critical impact parameter, and then establishing the eikonal correspondence between the shadow and the quasinormal frequency. The resulting analytical relationship is compared with the results obtained by the WKB approximation for the quasinormal modes, providing a consistency check between the geodesic and perturbative descriptions. 
Finally, we compare the predicted shadow size with the observational constraints from Sgr A*. We use the geometric correspondence between shadow radius and quasinormal modes to investigate how the observational constraints on the shadow translate into limitations for the GUP-corrected quasinormal response. This provides a unified framework for evaluating the feasibility of Schwarzschild–Tangherlini  geometry with GUP corrections and, in particular, for determining whether configurations in higher dimensions can remain compatible with current observations in strong field regimes.

The paper is organized as follows. In Sec.~\ref{s1} we introduce the Schwarzschild-Tangherlini black hole, considering quantum corrections arising from the Generalized Uncertainty Principle (GUP), and derive the radial equation for the model. In Sec~\ref{s2} we calculate the quasinormal modes using the WKB approximation for a Schwarzschild-Tangherlini black hole with GUP corrections and verify the results via time-domain evolution. In Sec.~\ref{s3} we present the analysis to calculate our model's shadow, establishing the correlation with the quasinormal modes, and finally, we constrain the quadratic correction parameter using EHT data for the shadow of Sgr A*. In Sec.~\ref{conc} we make our final considerations.

%%%%%%%%%%%%%%%%%%%%%%%%%%%%%%%%%%%%%%%%%%%%%%%%%
\section{Black Hole in extra space dimensions with GUP}
\label{s1}
The Schwarzschild-Tangherlini spacetime is a solution to Einstein's field equations in $(d+1)$ spacetime dimensions~\cite{Tangherlini:1963bw}. The exterior solution for a spherically symmetric and neutral black hole in a spacetime of extra dimensions is described by the metric
\begin{equation}
ds^2 = \left(1-\dfrac{2\mu}{r^{d-2}}\right)dt^2 - \left(1-\dfrac{2\mu}{r^{d-2}}\right)^{-1}dr^2 - r^2 d\Omega_{d-1}^2,
\label{metric1}
\end{equation}
where
\begin{eqnarray}
d\Omega_{d-1}^2 = d\theta_{d-2}^2 + \sin^2{\theta_{d-2}^2}\left[d\theta_{d-3}^2 + \sin^2\theta_{d-3}\left(... + \sin^2\theta_{2}\left(d\theta_{1}^2+\sin^2\theta_{1}d\phi^2\right)...\right) \right],
\end{eqnarray}
is the solid angle element in $(d-1)$ spatial dimensions, and $\Omega_{d-1}$ is the volume of the unit sphere $(d-1)$:  $\Omega_{d-1} = {2\pi^{d/2}}/{\Gamma(d/2)}$.

The parameter $\mu$ is a constant and is related to the black hole mass $M$ as follows:
\begin{equation}
\mu = \dfrac{16\pi M}{2(d - 1)\Omega_{d-1}}=\dfrac{8 M\Gamma(d/2)}{2(d-1)\pi^{(d-2)/2}},
\end{equation}
Note that by setting $d=3$ in \eqref{metric1} the metric reduces to the ordinary Schwarzschild spacetime. 
In addition, if the mass parameter $\mu$ is less than zero, we will have a non-physical situation. However, for positive values of this parameter, the radius of the horizon can be obtained by taking $g_{tt}(r_{h})=0$
\begin{equation}
\label{rh}
r_{h}^{d-2}=2\mu = \dfrac{16\pi M}{(d-1)\Omega_{d-1}}
=\dfrac{8 M\Gamma(d/2)}{(d-1)\pi^{(d-2)/2}}.
\end{equation}
In this work, we adopt GUP-corrected Schwarzschild-Tangherlini metric derived in~\cite{Anacleto:2023ntm}.
The modification of the Heisenberg uncertainty relation is written in the form
\begin{eqnarray}
    \Delta x \Delta p \geq \dfrac{\hbar}{2} \left[1 - \dfrac{\alpha_{0}\ell_{p}}{\hbar}\Delta p + \dfrac{\beta_{0}\ell_{p}^{2}}{\hbar^{2}} (\Delta p)^2\right],
\end{eqnarray}
%where $\alpha$ and $\beta$ are the linear and quadratic parameters of the GUP, respectively.
%These modifications leads to a measurable minimum length $\Delta x_{min}\sim \alpha$ and affects the structure of the black hole horizon.
where $\alpha_{0}$ and $\beta_{0}$ are dimensionless parameters characterizing the linear and quadratic GUP corrections, respectively, and $\ell_{p}$ denotes a fundamental length scale.
The modified uncertainty relation introduces a characteristic minimal length scale, and consequently modifies the near horizon structure of black holes.
 
Following the procedure of refs~\cite{Anacleto:2020lel,Anacleto:2021qoe} the GUP correction is implemented via an effective modification of the horizon scale:
\begin{eqnarray}
r_{hgup}^{d-2}=2\mu_{gup}=r_{h}^{d-2}\left[1 - \dfrac{4\alpha}{r_{h}^{d-2}} + \dfrac{16\beta}{r_{h}^{2(d-2)}} \right],
\label{rgup_rh}
\end{eqnarray}
where $\alpha$ and $\beta$ denote effective GUP parameters entering the higher dimensional black hole geometry. In $D=d+1$ dimensions, adopting $G_{D}=c=\hbar=1$, the mass parameter has dimension $[M]=L^{d-2}$, where $L$ denotes length.
Since $[r_{h}]=L$, dimensional consistency of Eq.~\eqref{rgup_rh} requires $[\alpha]=L^{d-2}$ and $[\beta]=L^{2(d-2)}$.

In terms of mass $M$, Eq.\eqref{rgup_rh} can be written in the form
\begin{eqnarray}
r_{hgup}^{d-2}=\dfrac{8M\Gamma(d/2)}{(d-1)\pi^{(d-2)/2}}\left[1 - \dfrac{(d-1)\pi^{(d-2)/2}\bar{\alpha} }{2\Gamma(d/2)} + \dfrac{(d-1)^{2}\pi^{(d-2)}\bar{\beta}}{4\Gamma(d/2)^{2}}\right],
\label{rgup_M}
\end{eqnarray}
here, for convenience, we introduce the dimensionless combinations $\bar{\alpha}=\dfrac{\alpha}{M}$ and $\bar{\beta}=\dfrac{\beta}{M^{2}}$.
%In this work, we adopt units $G_{D}=c=1$, in $D=d+1$ dimensions, which implies that the black hole mass has dimension $[M]=L^{d-2}$, where $L$ denotes length. Consequently, the GUP parameters $\alpha$ and $\beta$ must have dimensions $[\alpha]=L^{d-2}$ and $[\beta]=L^{2(d-2)}$ to ensure that the combinations $\alpha/r_{h}^{d-2}$ and $\beta/r_{h}^{2(d-2)}$ are dimensionless.

%-----------------
\subsection{Radial equation}
The Schwarzschild-Tangherlini metric with GUP corrections can be written in the form
\begin{eqnarray}
    ds^2 = B(r)dt^{2}- B(r)^{-1}dr^{2}-r^{2}d\Omega^{2}_{d-1},
    \label{metric2}
\end{eqnarray}
where $B(r) = 1 - \left(r_{hgup}/r\right)^{d-2}$ is the metric function. The dynamics of a scalar field is governed by the Klein-Gordon equation 
\begin{eqnarray}
\dfrac{1}{\sqrt{-g}}\partial_{\mu}\Big(\sqrt{-g}g^{\mu\nu}\partial_{\nu}\Psi\Big)=0 .
\label{eq_kleingordon}
\end{eqnarray}

Considering the background field described by the metric \eqref{metric2} and using the method of separation of variables, we define the appropriate ansatz for a spacetime of $(d+1)$ dimensions, we write:
\begin{eqnarray}
\Psi_{\omega l m}({\bf r},t)=\frac{\psi_{\omega l}(r)}{r^{(d-1)/2}} \, Y_{lm}(\Omega_{d-1}) \, e^{-i\omega t},
\label{ansatz_correct}
\end{eqnarray}
where $\omega$ is the frequency, $Y_{lm}(\Omega_{d-1})$ are the spherical harmonics in dimensions $(d-1)$, and the factor $r^{-(d-1)/2}$ is introduced to obtain a Schr\"odinger-type equation.

Substituting the ansatz \eqref{ansatz_correct} into the Klein-Gordon equation \eqref{eq_kleingordon} 
\begin{eqnarray}
\label{eqrad}
B(r)\dfrac{d}{dr}\left[B(r)\dfrac{d\psi_{\omega l}(r)}{dr} \right] +\left[ \omega^2 -V_{\text{eff}} \right]\psi_{\omega l}(r)=0,
\end{eqnarray}
The effective potential $V_{\text{eff}}(r)$ is given by:
\begin{eqnarray}
V_{\text{eff}}(r) = \frac{B(r)}{r^2} \Bigg[ l(l+d-2) + \frac{(d-3)(d-1)}{4} +  \frac{(d-1)^2}{4}\left(\frac{r_{hgup}}{r}\right)^{d-2}\Bigg].
\label{veff}
\end{eqnarray}
For the particular case of $d = 3$ (usual Schwarzschild black hole), the effective potential reduces to the well known form:
\begin{eqnarray}
V_{\text{eff}}^{(d=3)}(r) = \frac{B(r)}{r^2}\left[ l(l+1) + \frac{r_{hgup}}{r} \right].
\label{veff_d3}
\end{eqnarray}

Using the tortoise coordinate $dr_{*} = dr/B(r)$, we obtain the radial equation \eqref{eqrad} in the form of a one-dimensional Schr\"odinger equation
\begin{eqnarray}
\dfrac{d^2\psi_{\omega l}}{dr_{*}^2}  +\left[ \omega^2 - V_{\text{eff}}(r_{*}) \right]\psi_{\omega l}=0.
\label{eqRtot_correct}
\end{eqnarray}
The tortoise coordinate can be determined in terms of a hypergeometric function as follows
\begin{eqnarray}
    r_{*}=r-rF\left(1,\dfrac{1}{d-2};\dfrac{d-1}{d-2};\dfrac{r^{d-2}}{r_{hgup}^{d-2}}\right).
\end{eqnarray}
 
\section{Quasinormal modes of higher-dimensional black hole}
\label{s2}
Quasinormal modes are solutions to the perturbation equations that satisfy specific boundary conditions: purely ingoing waves at the event horizon and purely outgoing waves at spatial infinity~\cite{Berti:2009kk}. In terms of the tortoise coordinate $r_{*}$, defined in Section~\ref{s1}, these boundary conditions can be written as
\begin{eqnarray}
\mathcal{\psi}_{\omega l}  \sim e^{\pm i\omega r_{*}}, \qquad (r_{*} \rightarrow  \pm \infty),
\label{condQNM}
\end{eqnarray}
where the upper sign corresponds to the limit $r_{*} \rightarrow + \infty$ while the lower sign corresponds to $r_{*} \rightarrow -\infty$.
The quasinormal frequencies $\omega_{n}$ that satisfy these conditions form a discrete spectrum, indexed by the number of overtones $n = 0, 1, 2, \dots$. These frequencies are complex, with the real part, $\mathrm{Re}(\omega)$, describing the oscillation frequency, while the imaginary part, $\mathrm{Im}(\omega)<0$, controls the damping of the mode.
The study of quasinormal modes in higher-dimensional black holes has been a subject of research for several years, providing insights into the stability and dynamical properties of gravitational configurations~\cite{Cardoso:2003vt, Konoplya:2011qq, Vieira:2023ylz}. Recent studies have investigated the Schwarzschild-Tangherlini geometry, focusing on the accuracy and convergence of quasinormal mode calculations in five dimensions~\cite{Han:2025cal, Matyjasek:2021xfg}.

\subsection{WKB approximation}
To compute the quasinormal spectrum, we will use the WKB approximation, a technique widely used in the literature for effective potentials with a single barrier, as is the case of  $V_{\text{eff}}$ in \eqref{veff}. A review of the higher-order WKB method can be found in~\citep{Konoplya:2019hlu, Matyjasek:2026yiu}. We will use the sixth-order WKB formula introduced by Konoplya~\citep{Konoplya:2003ii}, which can be written in the form
\begin{eqnarray}
\dfrac{i\left(\omega_{n}^{2} - V_{0}\right)}{\sqrt{-2V''_{0}}} -\sum_{j=2}^{6} \Omega_{j} = n + \dfrac{1}{2},
\end{eqnarray}
where $V_{0}$ is the value of the effective potential at its maximum point (in tortoise coordinate $r_{*}$), $V''_{0} \equiv d^2V/dr_{*}^2$ evaluated at the same point, and $\Omega_{j}$ are the higher-order correction terms.

Tables \ref{tab1} and \ref{tab2} summarize the quasinormal frequencies for the fundamental mode and the first two overtones for $l=1$ and $l=2$. In all cases, we define a dimensionless natural frequency of the form $\hat{\omega}=\omega M^{1/(d-2)}$ and consider four combinations of GUP parameters: ($\bar{\alpha}$,$\bar{\beta}$) = ($0,0$), ($0,0.03$),($0.04,0$) and ($0.04,0.03$). The results show a dependence of the quasinormal spectrum on both the dimensionality of spacetime and the linear and quadratic GUP parameters. 
For the uncorrected case, both the real and imaginary parts of the quasinormal frequency increase as the number of dimensions grows for both $l=1$ and $l=2$. The results indicate that black holes in higher dimensions exhibit higher oscillation frequencies and a faster damping rate within the considered parameter range. For the linear correction of the GUP, $\bar{\alpha}$, the real part of the frequency increases in all dimensions compared to the uncorrected case. Simultaneously, the magnitude of the imaginary part increases, indicating faster damping of the perturbation. In contrast, the quadratic correction controlled by $\bar{\beta}$ produces a reduction in both the real part and the magnitude of the imaginary part of the quasinormal frequencies, that is, a reduction in the damping rate. When both corrections are present, we have a competition between the linear and quadratic terms of the GUP. 
The dependence on the angular quantum number can be seen in the two tables. For a fixed dimension and GUP parameters, increasing $l$ raises the real part of the quasinormal frequency, as expected from the angular contribution $l(l+d-2)$ to the effective potential. The comparison between $l=1$ and $l=2$ provides a useful consistency check regarding the dependence of the quasinormal spectrum on the angular sector.

\begin{table}[h!]
	\begin{center}
	\caption{\footnotesize{Quasinormal frequencies for $l = 1$.}} 
	\label{tab1}
\begin{tabular}{|c||c||c|c|c|}
\hline
 \multicolumn{5} {|c|}{ $d=3$ }  \\
\hline
 $\bar{\alpha}$ & $\bar{\beta}$   &  $\hat{\omega}_{0}$ &   $\hat{\omega}_{1}$ &  $\hat{\omega}_{2}$ \\
 \hline
\multirow{2}{*}{$ 0.00 $} 
   & 0.00  & 0.292910 - 0.097762i & 0.264471 - 0.306518i  & 0.231014 - 0.542166i  \\
   & 0.03  & 0.261526 - 0.087287i & 0.236135 - 0.273677i  & 0.206263 - 0.484076i  \\
   \hline
\multirow{2}{*}{$ 0.04 $}
   & 0.00  & 0.318380 - 0.106263i  & 0.287469 - 0.333172i  & 0.251102 - 0.589310i  \\
   & 0.03  & 0.281644 - 0.094002i  & 0.254299 - 0.294729i  & 0.222129 - 0.521313i  \\
 \hline
 \hline
 \multicolumn{5} {|c|}{ $d=4$ }  \\
\hline
$\bar{\alpha}$ & $\bar{\beta}$   &  $\hat{\omega}_{0}$ &   $\hat{\omega}_{1}$ &  $\hat{\omega}_{2}$  \\
 \hline
\multirow{2}{*}{$ 0.00 $} 
   & 0.00  & 1.101080 - 0.396434i & 0.928592 - 1.266000i  & 0.69154 - 2.33684i  \\
   & 0.03  & 0.853011 - 0.307120i & 0.719385 - 0.980776i  & 0.53574 - 1.81037i  \\
   \hline
\multirow{2}{*}{$ 0.04 $}
   & 0.00  & 1.222290 - 0.440074i  & 1.030810 - 1.405360i  & 0.767665 - 2.59409i   \\
   & 0.03  & 0.905784 - 0.326120i  & 0.763891 - 1.041450i  & 0.568884 - 1.92237i   \\
 \hline
 \hline
 \multicolumn{5} {|c|}{ $d=5$ }  \\
\hline
 $\bar{\alpha}$ & $\bar{\beta}$   &  $\hat{\omega}_{0}$ &   $\hat{\omega}_{1}$ &  $\hat{\omega}_{2}$  \\
 \hline
\multirow{2}{*}{$ 0.00 $} 
   & 0.00  & 1.84019 - 0.666898i & 1.45778 - 2.15343i  & 0.79054 - 4.15285i  \\
   & 0.03  & 1.26130 - 0.457104i & 0.99919 - 1.47600i  & 0.54185 - 2.84644i  \\
   \hline
\multirow{2}{*}{$ 0.04 $}
   & 0.00  & 2.10836 - 0.764084i  & 1.67022 - 2.46724i  & 0.905744 - 4.75804i   \\
   & 0.03  & 1.31023 - 0.474837i  & 1.03795 - 1.53326i  & 0.562871 - 2.95687i   \\
 \hline
 \hline
 \multicolumn{5} {|c|}{ $d=6$ }  \\
\hline
 $\bar{\alpha}$ & $\bar{\beta}$   &  $\hat{\omega}_{0}$ &   $\hat{\omega}_{1}$ &  $\hat{\omega}_{2}$ \\
 \hline
\multirow{2}{*}{$ 0.00 $} 
   & 0.00  & 2.45993 - 0.891478i & 1.83496 - 2.89591i  & 0.458929 - 5.69388i  \\
   & 0.03  & 1.60153 - 0.580395i & 1.19465 - 1.88538i  & 0.298785 - 3.70699i  \\
   \hline
\multirow{2}{*}{$ 0.04 $}
   & 0.00  & 2.91591 - 1.056720i  & 2.17509 - 3.43271i  & 0.543996 - 6.74930i  \\
   & 0.03  & 1.63914 - 0.594024i  & 1.22270 - 1.92965i  & 0.305800 - 3.79403i  \\
 \hline
 \end{tabular}
\end{center}
\end{table}

\begin{table}[h!]
	\begin{center}
	\caption{\footnotesize{Quasinormal frequencies for $l = 2$.}} 
	\label{tab2}
\begin{tabular}{|c||c||c|c|c|}
\hline
 \multicolumn{5} {|c|}{ $d=3$ }  \\
\hline
 $\bar{\alpha}$ & $\bar{\beta}$   &  $\hat{\omega}_{0}$ &   $\hat{\omega}_{1}$ &  $\hat{\omega}_{2}$ \\
 \hline
\multirow{2}{*}{$ 0.00 $} 
   & 0.00  & 0.483642 - 0.0967661i & 0.463847 - 0.295627i  & 0.430386 - 0.508700i \\
   & 0.04  & 0.431823 - 0.0863983i & 0.414149 - 0.263953i  & 0.384273 - 0.454196i \\
   \hline
\multirow{2}{*}{$ 0.03 $}
   & 0.00  & 0.525698 - 0.105181i  & 0.504181 - 0.321334i  & 0.467811 - 0.552935i  \\
   & 0.03  & 0.465040 - 0.093044i  & 0.446006 - 0.284257i  & 0.413832 - 0.489135i  \\
 \hline
 \hline
\multicolumn{5} {|c|}{ $d=4$ }  \\
\hline
 $\bar{\alpha}$ & $\bar{\beta}$   &  $\hat{\omega}_{0}$ &   $\hat{\omega}_{1}$ &  $\hat{\omega}_{2}$\\
 \hline
\multirow{2}{*}{$ 0.00 $} 
   & 0.00  & 1.63950 - 0.388248i & 1.51141 - 1.19977i  & 1.287710 - 2.11467i \\
   & 0.03  & 1.27013 - 0.300778i & 1.17090 - 0.92947i  & 0.997599 - 1.63825i \\
   \hline
\multirow{2}{*}{$ 0.04 $}
   & 0.00  & 1.81998 - 0.430986i  & 1.67779 - 1.33184i  & 1.42947 - 2.34746i  \\
   & 0.03  & 1.34871 - 0.319386i  & 1.24334 - 0.98697i  & 1.05932 - 1.7396i  \\
 \hline
 \hline
 \multicolumn{5} {|c|}{ $d=5$ }  \\
\hline
 $\bar{\alpha}$ & $\bar{\beta}$   &  $\hat{\omega}_{0}$ &   $\hat{\omega}_{1}$ &  $\hat{\omega}_{2}$  \\
 \hline
\multirow{2}{*}{$ 0.00 $} 
   & 0.00  & 2.57351 - 0.642702i & 2.28370 - 2.00043i  & 1.73135 - 3.60597i  \\
   & 0.03  & 1.76393 - 0.440520i & 1.56529 - 1.37113i  & 1.18670 - 2.47160i  \\
   \hline
\multirow{2}{*}{$ 0.04 $}
   & 0.00  & 2.94855 - 0.736363i  & 2.61651 - 2.29195i  & 1.98366 - 4.13147i \\
   & 0.03  & 1.83236 - 0.457609i  & 1.62602 - 1.42432i  & 1.23274 - 2.56748i \\
 \hline
 \hline
 \multicolumn{5} {|c|}{ $d=6$ }  \\
\hline
 $\bar{\alpha}$ & $\bar{\beta}$   &  $\hat{\omega}_{0}$ &   $\hat{\omega}_{1}$ &  $\hat{\omega}_{2}$ \\
 \hline
\multirow{2}{*}{$ 0.00 $} 
   & 0.00  & 3.30920 - 0.837986i & 2.82722 - 2.62485i  & 1.82654 - 4.82614i  \\
   & 0.03  & 2.15445 - 0.545569i & 1.84066 - 1.70891i  & 1.18917 - 3.14205i  \\
   \hline
\multirow{2}{*}{$ 0.04 $}
   & 0.00  & 3.92260 - 0.993317i  & 3.35128 - 3.11140i  & 2.16511 - 5.72073i  \\
   & 0.03  & 2.20504 - 0.558380i  & 1.88388 - 1.74903i  & 1.21709 - 3.21583i  \\
 \hline
 \end{tabular}
\end{center}
\end{table}

\begin{figure}[htbp]
 \centering
  \subfigure[]{\includegraphics[scale=.4]{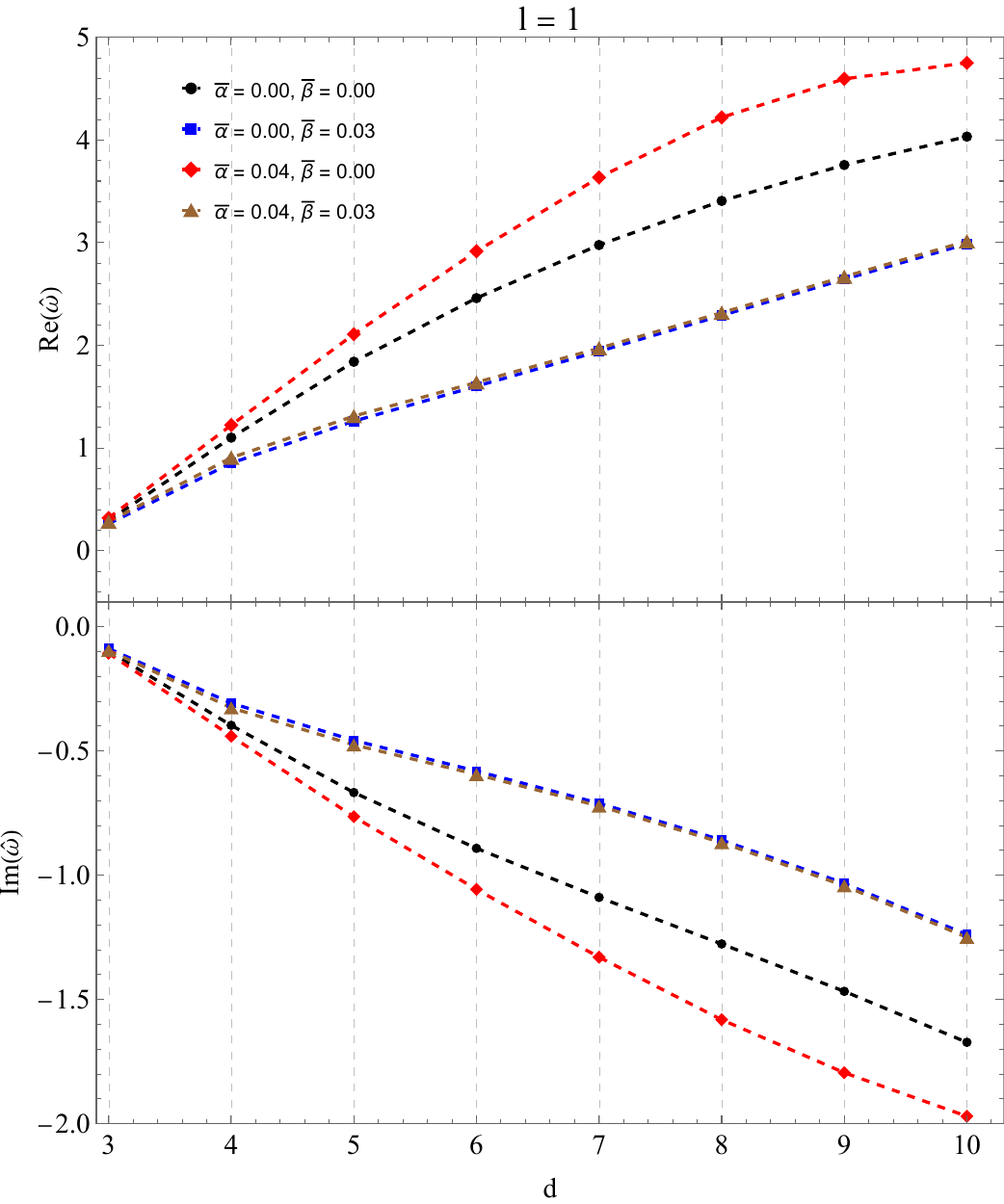}\label{QNMdl1}}
  \quad
   \subfigure[]{\includegraphics[scale=.4]{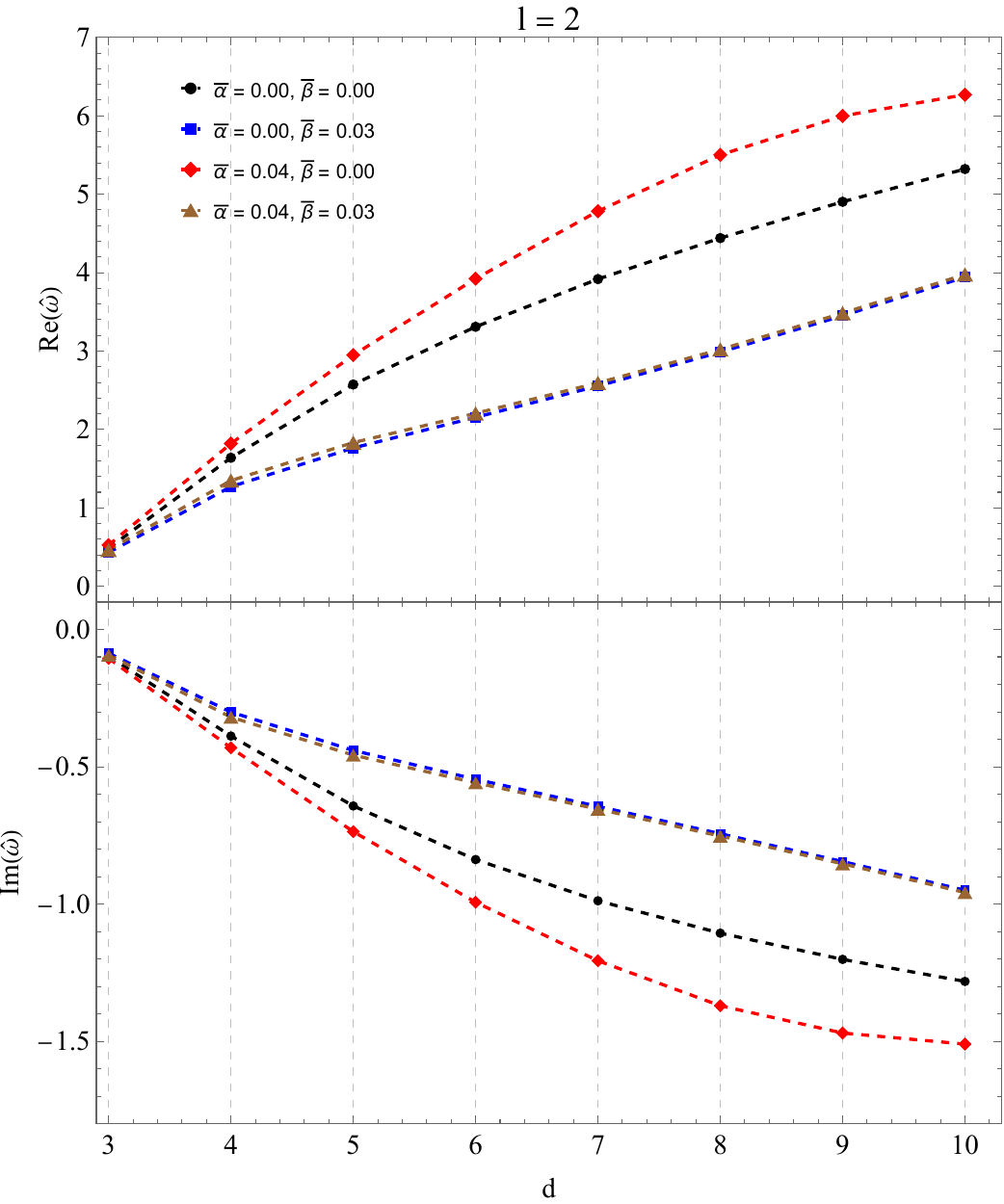}\label{QNMdl2}}
  \caption{\footnotesize{Dependence of the quasinormal frequencies on the dimensionality $d$ for (a) $l=1$ and (b) $l=2$, considering different GUP correction parameters.}}
  \label{QNMd}
 \end{figure}

Figure~\ref{QNMd} shows the dependence of the real and imaginary parts of the quasinormal frequency on the spacetime dimensionality for $l=1$ and $l=2$. The different curves correspond to the Schwarzschild-Tangherlini case  and to cases with specific values of $\bar{\alpha}$ and $\bar{\beta}$. This allows us to identify how extra dimensions and corrections affect the oscillation and damping scales. 

For geometry without corrections, both the real part and the magnitude of the imaginary part increase as the number of dimensions rises, meaning higher dimensions exhibit higher oscillation frequencies and greater damping rates, indicating that perturbations decay more rapidly as the dimension increases. This behavior is observed for both $l=1$ and $l=2$, although the absolute frequency values are higher for $l=2$. The inclusion of GUP corrections modifies this dimensional dependence. Linear correction $\bar{\alpha}$ leads simultaneously to faster oscillations and a shorter damping timescale; conversely, quadratic contribution $\bar{\beta}$ results in a reduction of both quantities in the cases considered. Thus, the two GUP parameters exert qualitatively distinct effects on the quasinormal spectrum.

\begin{figure}[!htb]
 \centering
 \subfigure[]{\includegraphics[scale=0.42]{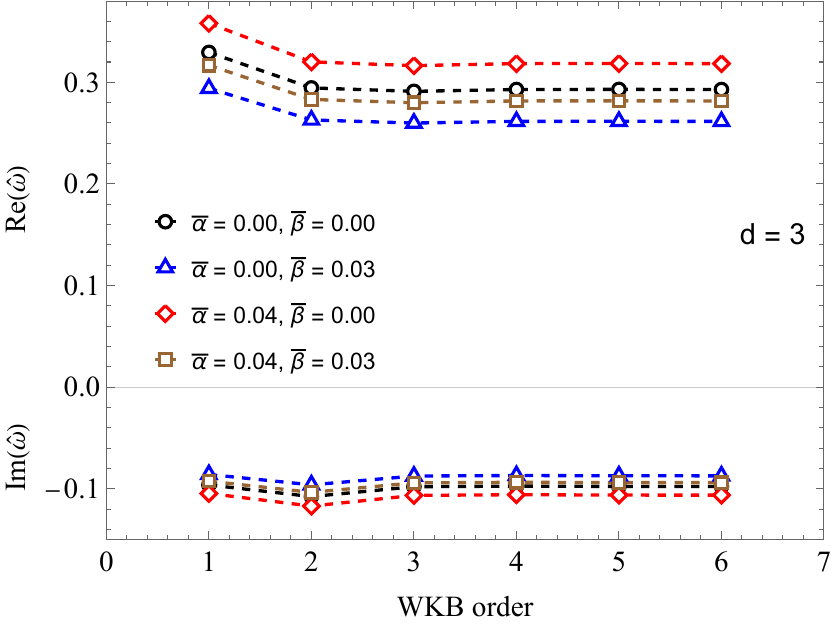}\label{OrdWKBd3}}
\quad
 \subfigure[]{\includegraphics[scale=0.42]{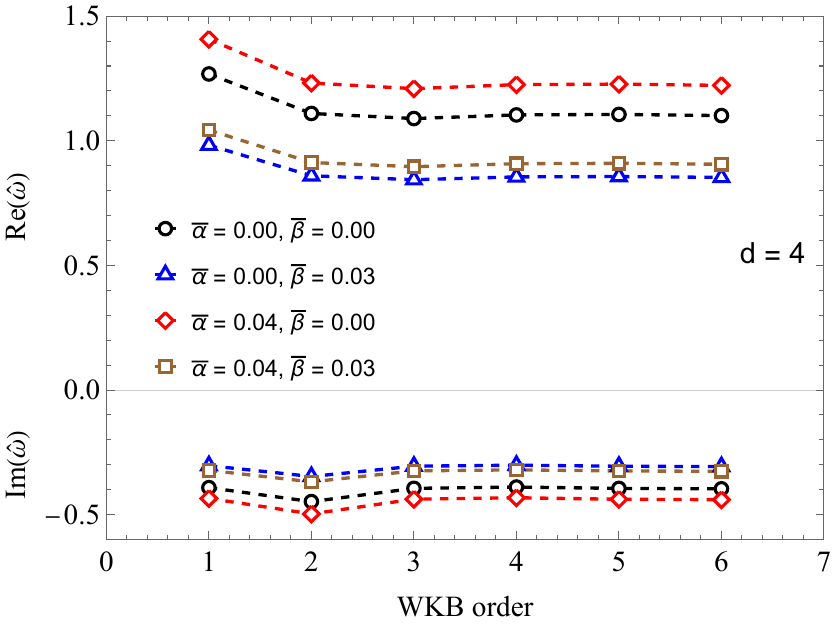}\label{OrdWKBd4}}
\quad
\subfigure[]{\includegraphics[scale=0.42]{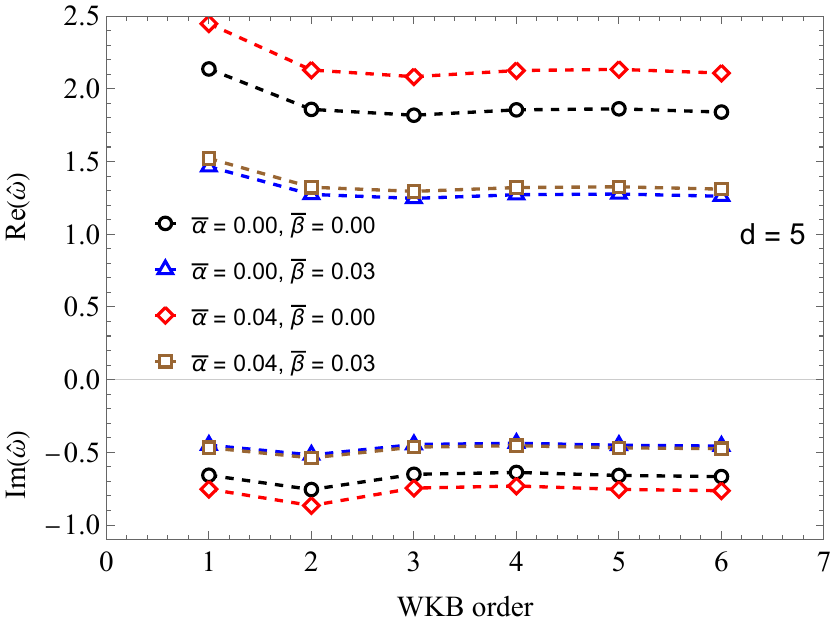}\label{OrdWKBd5}}
\quad
\subfigure[]{\includegraphics[scale=0.42]{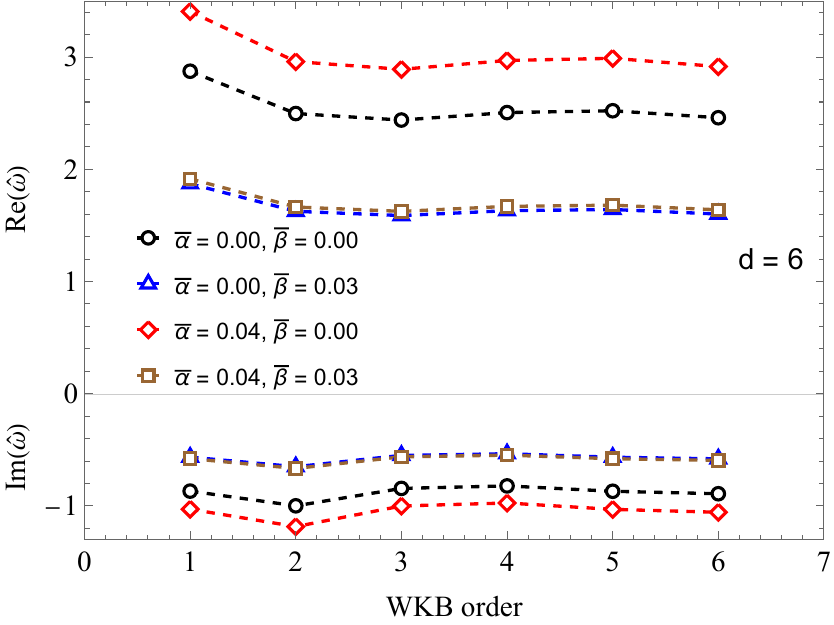}\label{OrdWKBd6}}
 \caption{\footnotesize{Behavior of the real and imaginary parts of the quasinormal frequencies as a function of the WKB expansion order (from second to sixth order) for (a) $d=3$, (b) $d=4$, (c) $d=5$, and (d) $d=6$. The four curves correspond to different combinations of GUP.}}
  \label{EvOrd}
\end{figure}

Figure~\ref{EvOrd} shows the behavior of the real and imaginary parts of the quasinormal frequencies as a function of the WKB order ranging from the second to the sixth order for $d=3$, $d=4$, $d=5$, and $d=6$. As the WKB order increases, we obtain a sequence of approximations whose stability can be used to test the reliability of the resulting quasinormal frequencies, although the degree of convergence depends on the spacetime dimension, the overtone number, and the GUP parameters. The behavior observed in the figure indicates that the WKB expansion is most reliable for modes where the stabilization of successive WKB orders provides an internal consistency check of the approximation.

\subsection{Time-domain}
We can examine the role of quasinormal modes in a time-dependent scattering process, that is, investigate the evolution of scalar perturbations in the time-domain. It is known that a perturbation in Schwarzschild spacetime evolves in three phases: an initial transient phase depending on the initial conditions; a ringdown-dominated phase, in which the primary contribution comes from quasinormal modes; and a third phase characterized by an approximately polynomial decay. During the ringdown phase, the damped oscillations have frequencies that depend solely on the parameters characterizing the black hole.
Rewriting the wave equation \eqref{eqRtot_correct} without imposing the stationary ansatz $\psi \sim R(r)e^{-i\omega t}$, we obtain the following form.
\begin{eqnarray}
\label{eqradtemp}
\dfrac{\partial^{2}\psi}{\partial t^{2}} - \dfrac{\partial^2\psi}{\partial r_{*}^2} + V_{eff} \psi = 0.
\end{eqnarray}

The technique for integrating the wave equation above in the time-domain was developed by Gundlach and collaborators \cite{Gundlach:1993tn}. By expressing the wave equation in terms of the light-cone coordinates $u = t - r_{*}$ and $v = t + r_{*}$ the wave equation can be rewritten as
\begin{equation}
\left(4\dfrac{\partial^{2}}{\partial u \partial v} + V(u,v)\right)\psi(u,v) = 0.
\label{eqtime}
\end{equation}

Equation \eqref{eqtime} can be integrated numerically using a finite-difference scheme. Applying a Taylor expansion to the wave function around the point $(u,v)$, we obtain
\begin{eqnarray}
\psi(u+h,v+h) &=& -\psi(u,v) + \psi(u+h,v) + \psi(u,v+h)\nonumber \\
& &-\dfrac{h^{2}}{8}V(u,v)\left[\psi(u+h,v) + \psi(u,v+h) \right] + \mathcal{O}\left(h^{2}\right)
\label{eqMDF}
\end{eqnarray}
where $h$ is the step size, assumed to be the same for $u$ and $v$. This scheme allows us to calculate the values of $\psi$ within a null grid—constructed upon two surfaces, $u = u_{0}$ and $v = v_{0}$, based on the specified initial data.
By choosing an initial Gaussian profile centered at $v = v_{c}$ with width $\sigma$ at $u = u_{0}$,
\begin{eqnarray}
\psi(u=u_{0}, v) = Ae^{-(v - v_{c})^{2} /2\sigma^{2}}, \qquad \psi(u, v=v_{0}) = \psi_{0}
\label{eqInitial}
\end{eqnarray}
Here, without loss of generality, we can assume $\psi_{0} = 0$. Once the initial values are defined, integration proceeds along lines of constant $u$ in the direction of increasing $v$. 
For our numerical implementation, we adopt the following parameters: a Gaussian centered at $v_{c} = 10$ with amplitude $A = 1$ and $u_{0} = 0$. The computational grid covers the interval $u, v \in [0,400]$ with $4000$ points, resulting in $h = 0.1$.

\begin{figure}[!htb]
 \centering
 \subfigure[]{\includegraphics[scale=0.36]{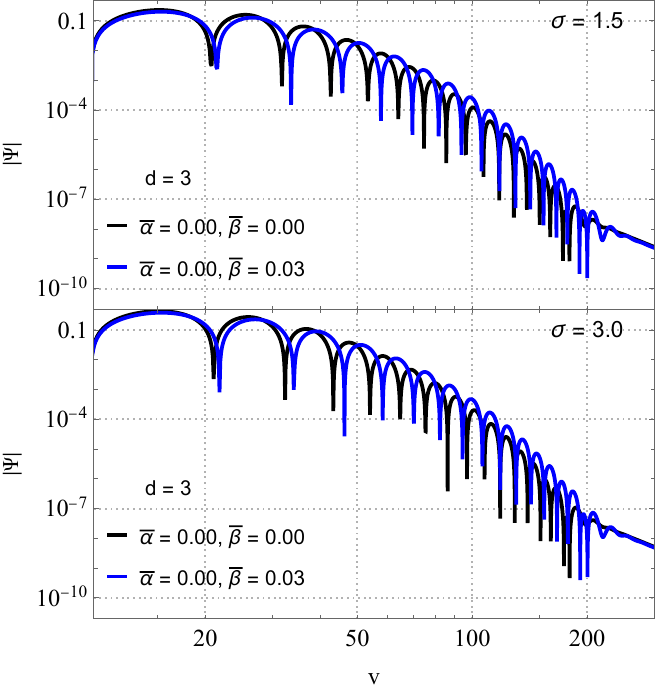}\label{QNMd3sig}}
 \quad
 \subfigure[]{\includegraphics[scale=0.36]{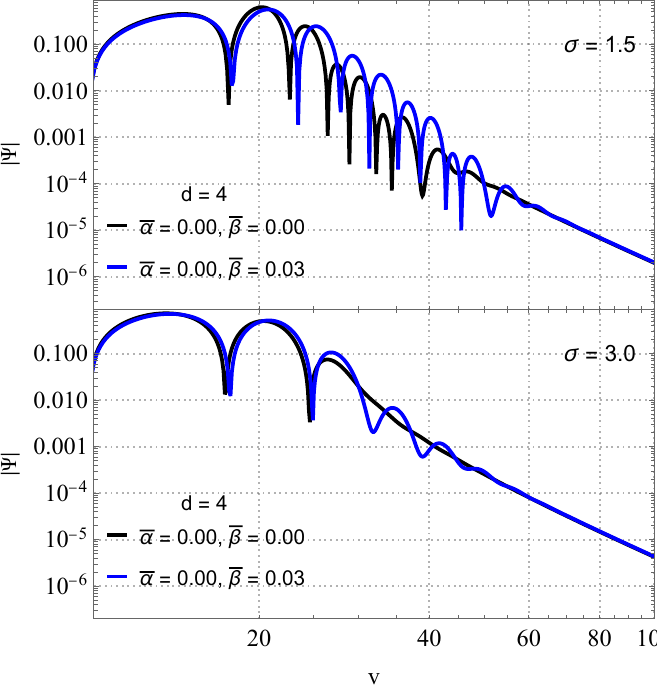}\label{QNMd4sig}}
  \quad
 \subfigure[]{\includegraphics[scale=0.36]{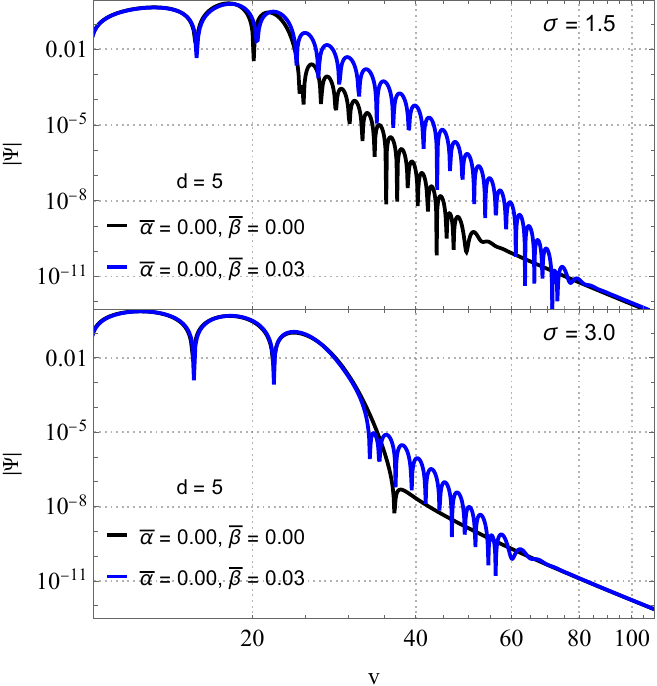}\label{QNMd5sig}}
  \quad
 \subfigure[]{\includegraphics[scale=0.36]{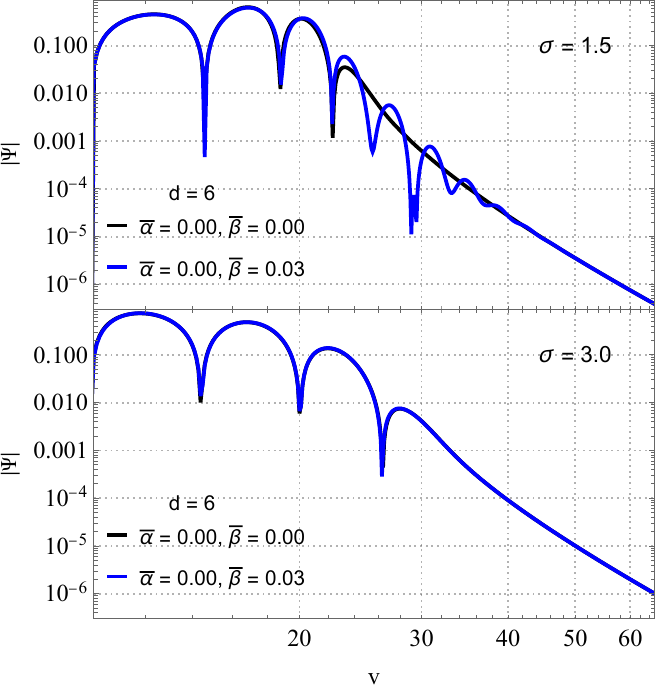}\label{QNMd6sig}}
\caption{\footnotesize{Test of the initial Gaussian width for the wave function, comparing the Schwarzschild-Tangherlini case with the GUP correction effect.}}
  \label{EvTempsig}
\end{figure}

In our case, we observe that higher dimensions are more sensitive to the width $\sigma$ of the initial Gaussian. To assess this sensitivity of the time-domain response to the initial perturbation, we compare the evolutions obtained for Gaussian widths of $\sigma=1.5$ and $\sigma=3$ in \eqref{eqInitial}, considering cases without GUP correction and with a quadratic correction of $\bar{\beta}=0.03$. As shown in Fig. \ref{EvTempsig}, for $d=3$, the corrected and uncorrected waveforms remain very similar for both choices of $\sigma$. However, for $d=4, 5$ and $6$, the narrower Gaussian ($\sigma=1.5$) produces a more pronounced oscillatory structure than the broader profile ($\sigma=3$). Thus, the dependence on $\sigma$ is particularly significant in higher dimensions, where the damping rate of quasinormal frequencies is higher. This causes the ringdown to be suppressed more rapidly, making the observable waveform increasingly sensitive to the excitation coefficients of the dominant modes. In particular, GUP-induced differences remain clearly visible for $d=6$ with $\sigma=1.5$ (Fig. \ref{QNMd6sig}), demonstrating that the visibility of the correction depends on the perturbation profile.  
This dependence should not be interpreted as an alteration of the quasinormal spectrum by the initial conditions. Rather, a narrower Gaussian contains a broader range of spectral components and can therefore excite the quasinormal response differently than a broader Gaussian profile.

\begin{figure}[!htb]
 \centering
 \subfigure[]{\includegraphics[scale=0.30]{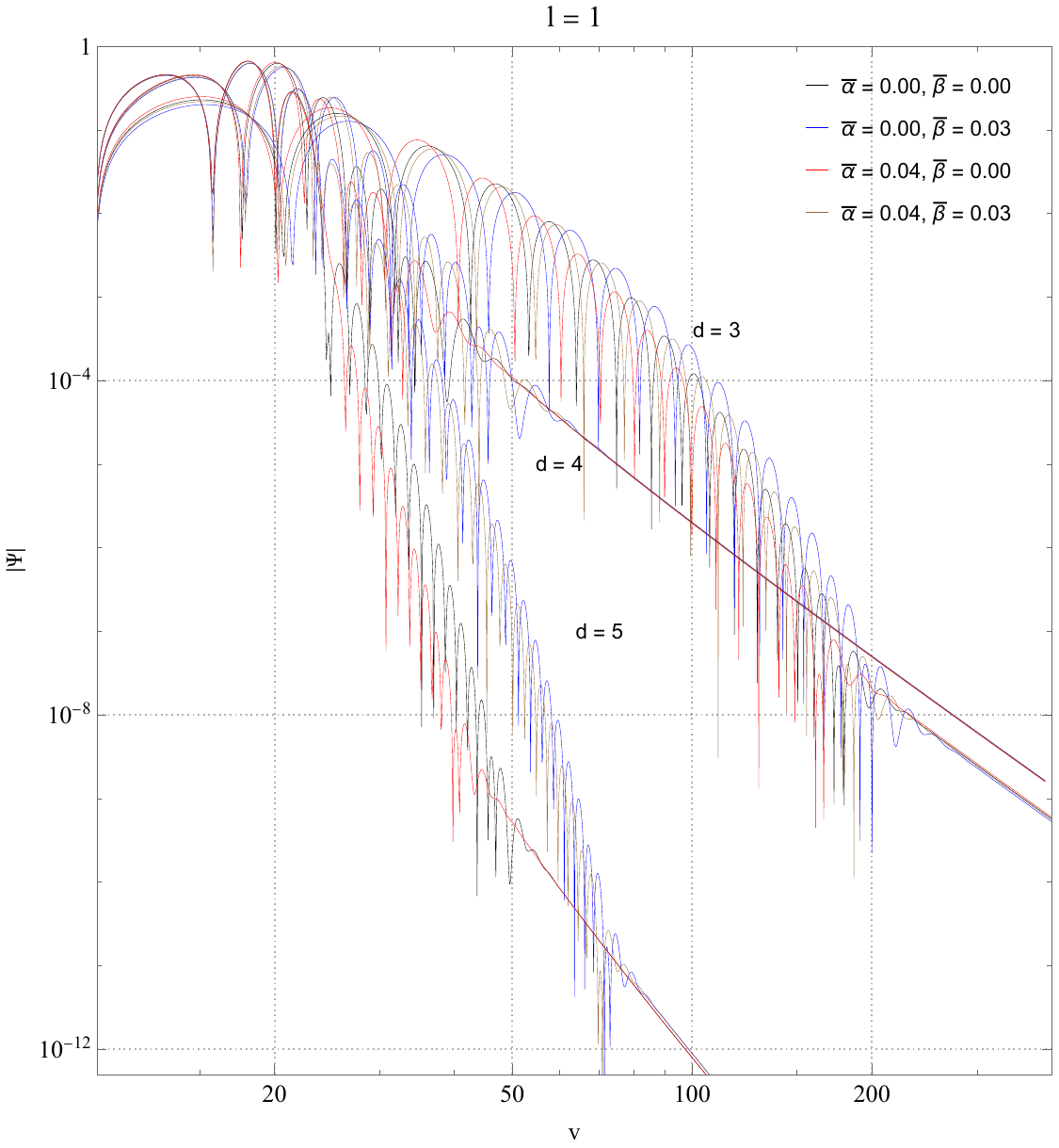}\label{QNM_MDFd3d4d5}}
 \qquad
 \subfigure[]{\includegraphics[scale=0.30]{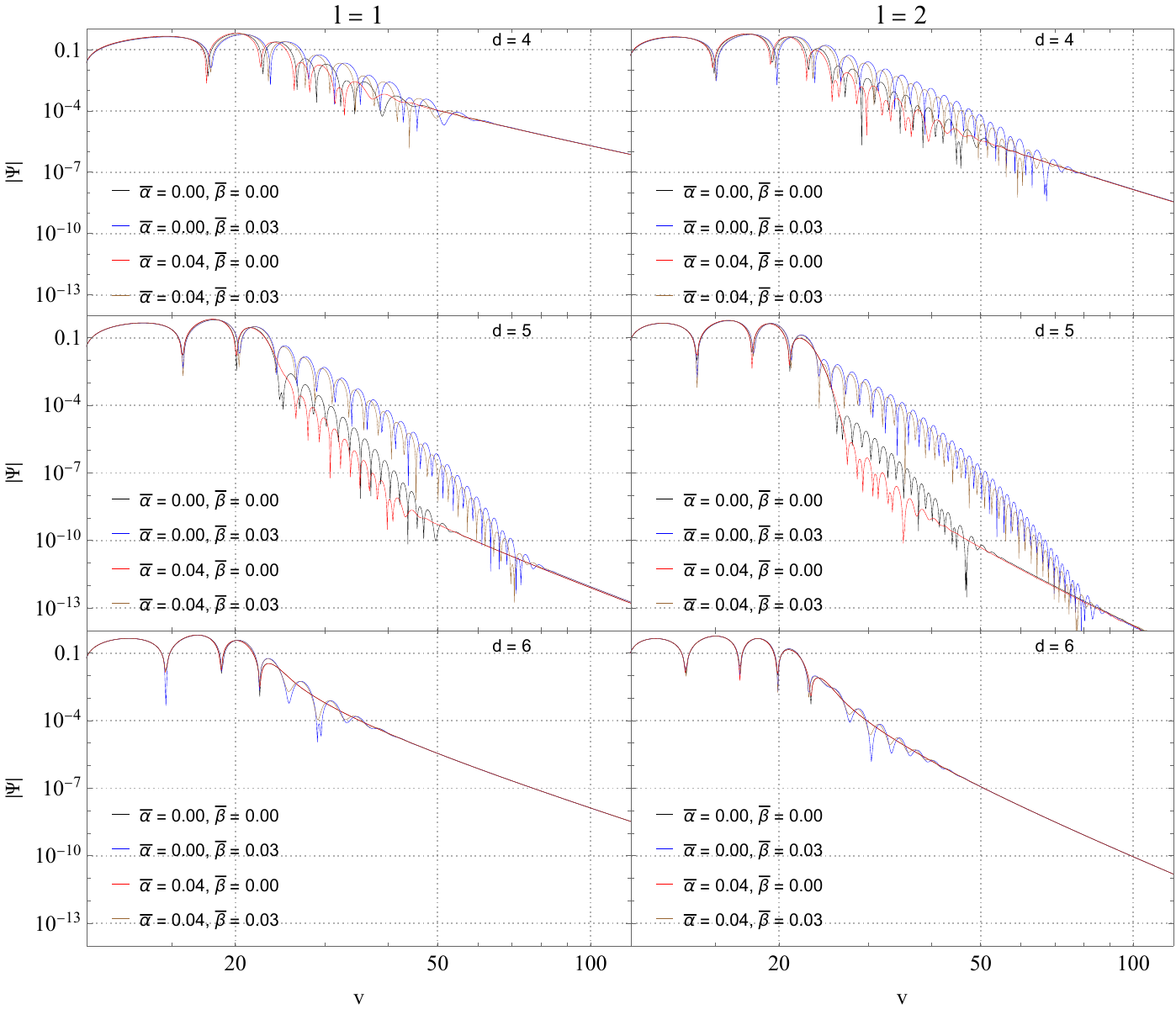}\label{QNM_MDFd4d5d6}}
\caption{\footnotesize{Log-log plots of the wave function, comparing the Schwarzschild-Tangherlini case with the effects of the $\bar{\alpha}$ and $\bar{\beta}$ parameters for each dimension. Left panel $d = 3, 4, 5$ with $l=1$, showing the effects of the $\bar{\alpha}$ and $\bar{\beta}$. Right panel, dimensions $d = 4, 5, 6$, comparing $l=1$ and $l=2$.}}
  \label{EvTemp}
\end{figure}

Figure~\ref{EvTemp} shows the time evolution of the scalar perturbation represented by the amplitude $|\Psi|$ on a logarithmic scale for spacetimes with $d=3$, $d=4$, $d=5$ and $d=6$. For each dimension, four combinations of GUP correction parameters used previously are considered. During the ringdown phase dominated by quasinormal modes, these different parameter choices alter the perturbation's decay profile, showing that GUP-induced corrections modify the dynamics of scalar perturbations.

Figure~\ref{QNM_MDFd3d4d5}, illustrates the behavior for $d=3, 4$ and $5$. Note that for $d>3$ without GUP corrections, damping is significantly faster, as linear or quadratic corrections, or both simultaneously, are introduced the wave's oscillation frequency and decay rate are modified. 
Figure~\ref{QNM_MDFd4d5d6}, compares dimensions $d=4$ through $d=6$, considering angular modes $l=1$ and $l=2$. Increasing the dimensionality alters both the decay rate and the oscillation structure, consistent with earlier WKB results. Furthermore, the differences between cases involving GUP corrections become particularly noticeable in higher dimensions, especially when contributions from the quadratic term are included.
The modes with $l=2$ exhibit decay behavior distinct from that of $l=1$, reflecting the contribution of the centrifugal term to the effective potential.

\section{Black Hole Shadow}
\label{s3}
The shadow of a black hole is determined by the unstable circular photon orbits. 
For this purpose, we will use the null geodesic equations. So, starting with the Lagrangian
\begin{equation}
\mathcal{L} \equiv \dfrac{1}{2}g_{\mu\nu}\dot{x}^{\mu}\dot{x}^{\nu},
\end{equation} 
and using the background metric~\eqref{metric1}, we obtain
\begin{equation}
2\mathcal{L} = B(r)\dot{t}^{2} - \dfrac{\dot{r}^{2}}{B(r)} - \sum_{i=1}^{d-2}r^{2}\prod_{n=1}^{i-1} \sin^2\theta_{n} \dot{\theta}_{d-2}^{2} - r^2 \prod_{i=1}^{d-2} \sin^{2}\theta_{i} \dot{\phi}_{d-2}^{2},
\label{elidot}
\end{equation} 
where the dot ``." denotes differentiation with respect to the affine parameter.

We are interested in the trajectory of a light ray in the described metric, which is spherically symmetric, so if we analyze in a plane, any ray of light that begins with a certain angle $ \theta $ must remain with the same angle. We will then consider the equatorial plane by setting $\theta =\pi/2$.

Thus, the conserved quantities associated with geodesic motion are energy $E$ and angular momentum $L$, given by

\begin{eqnarray}
E = B(r)\dot{t}, \qquad L = r^{2}\dot{\phi}.
\label{eqEL}
\end{eqnarray}
For the case of null geodesics $g_{\mu\nu}\dot{x}^{\mu}\dot{x}^{\nu} = 0$, we obtain 
\begin{equation}
\dot{r}^{2} = E^{2} - B(r)\dfrac{L^{2}}{r^{2}} \qquad \text{and} \qquad \ddot{r} = \dfrac{L^{2}}{2r^{3}}\left[2B(r)-rB'(r)\right] .
\label{eqEner}
\end{equation}
we find the critical radius $r_{c}$ and the critical impact parameter $b_{c}$, applying the following conditions $\dot{r}=0$ and $\ddot{r}=0$.
From the second derivative of $r$, we obtain the critical radius:
\begin{equation}
2B(r_{c})-B'(r_{c})r_{c}= 0 \quad \Rightarrow \quad r_{c}^{d-2} = \frac{d}{2} r_{hgup}^{d-2},
\label{photon_sphere}
\end{equation}
and from the first derivative of $r$, we obtain the critical impact parameter
\begin{equation}
b_c = \frac{r_{c}}{\sqrt{B(r_{c})}} = r_{c} \sqrt{\frac{d}{d-2}} = \left(\frac{d}{2}\right)^{\frac{1}{d-2}} r_{hgup} \sqrt{\frac{d}{d-2}},
\label{critical_impact}
\end{equation}
where the impact parameter is given by $b=L/E$.

The shadow of a black hole is related to the absorption of photons near the event horizon. This observable carries direct information about spacetime geometry and serves as one of the probes of strong-field gravity.
To systematically estimate the shadow size, we introduce the celestial coordinates ($\xi$, $\eta$) for an observer at infinity. These coordinates are defined as the apparent impact parameters of a light ray on the observer's plane of sight \citep{Vazquez:2003zm} 
\begin{eqnarray}
\xi & = & \lim\limits _{r_{o}\to\infty}\left[-r_{o}^{2}\sin\theta_{o}\dfrac{d\phi}{dr}\Bigr\rvert_{\theta=\theta_{o}}\right],\label{coodcelesta}\\
\eta & = & \lim\limits _{r_{o}\to\infty}\left[r_{o}^{2}\dfrac{d\theta}{dr}\Bigr\rvert_{\theta=\theta_{o}}\right],\label{coodcelestb}
\end{eqnarray}
where $\left(r_{o},\theta_{o}\right)$ is the position of the observer at infinity. Thus, for an observer on an equatorial plane $\theta_{o}=\pi/2$, we have the following relationship for the shadow radius 

\begin{equation}
R_{s} \equiv\sqrt{\xi^{2}+\eta^{2}}= \left(\frac{d}{2}\right)^{\frac{1}{d-2}} r_{hgup} \sqrt{\frac{d}{d-2}}.
\label{Rs_d}
\end{equation}

\begin{figure}[!htb]
 \centering
 \subfigure[]{\includegraphics[scale=0.4]{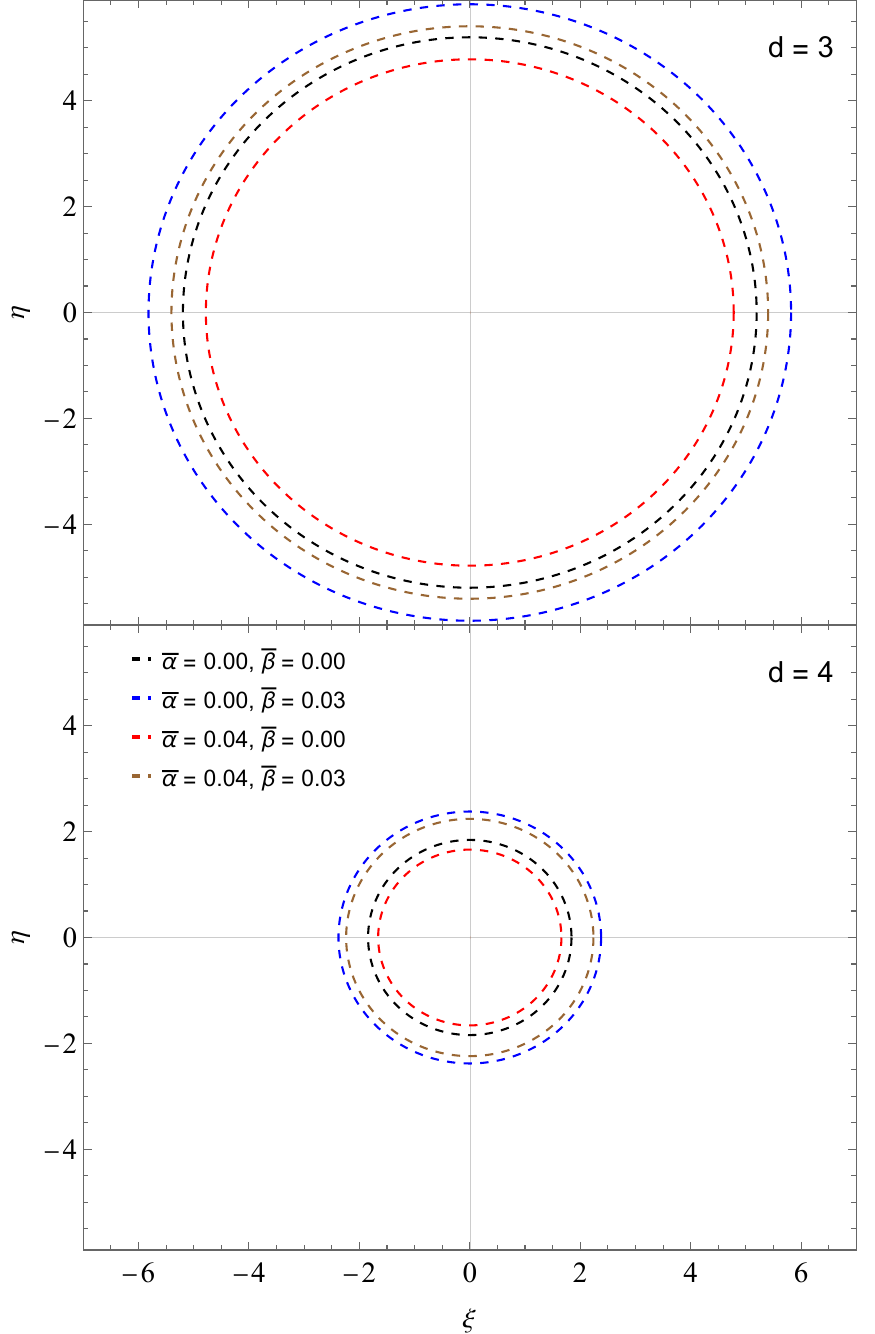}\label{Rsd3d4}}
 \qquad
 \subfigure[]{\includegraphics[scale=0.4]{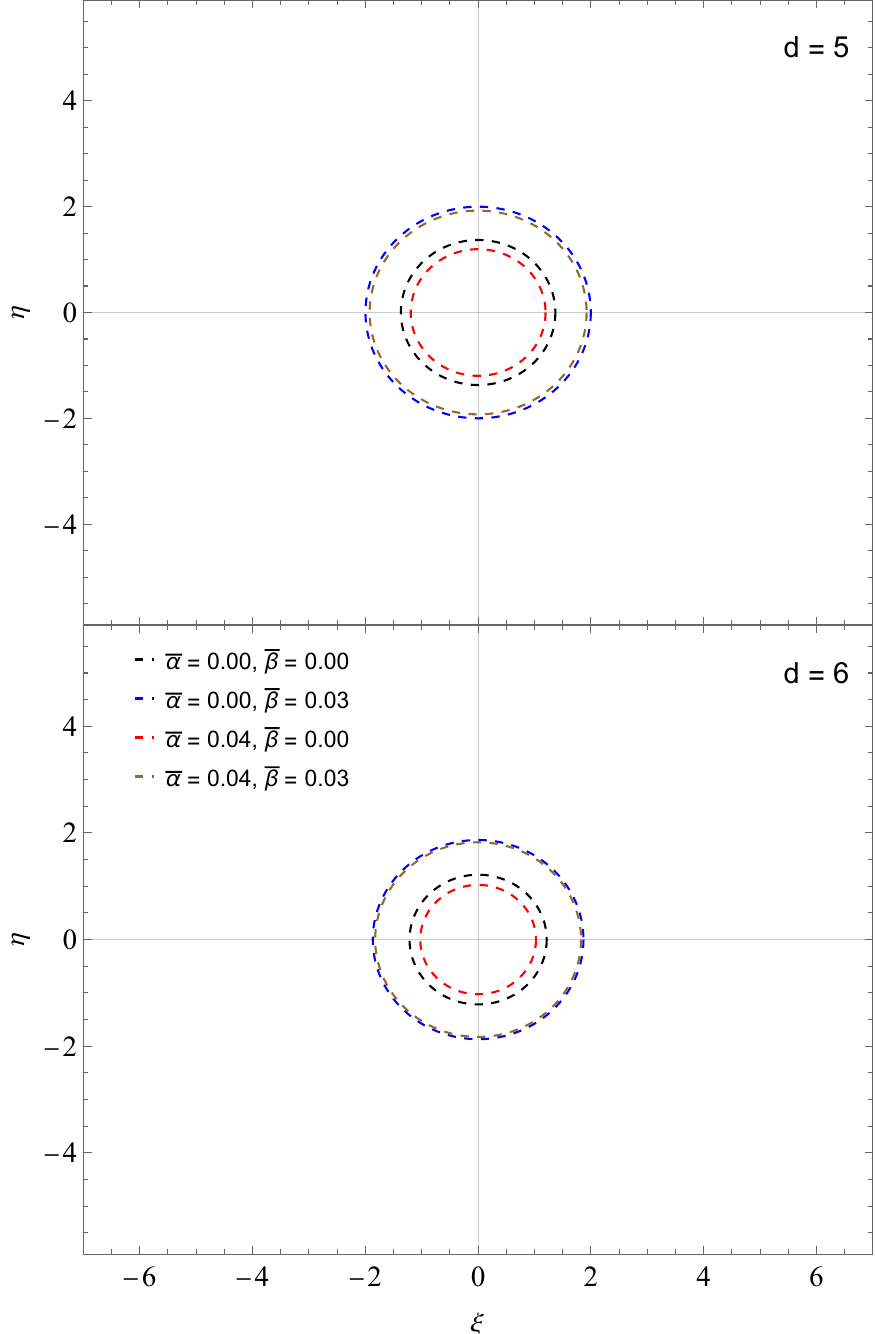}\label{Rsd5d6}}
 \caption{\footnotesize{We have the shadow radius represented by the dotted circles. Each panel corresponds to a distinct dimension, with the respective influence of the linear and quadratic GUP corrections.}}
  \label{Shadow}
\end{figure}

In terms of parameters $\bar{\alpha}$ and $\bar{\beta}$, using \eqref{rgup_M}, we obtain:
\begin{equation}
\mathcal{R}_{s}=\dfrac{R_{s}}{M^{1/(d-2)}} = \left(\frac{d}{2}\right)^{\frac{1}{d-2}} \sqrt{\frac{d}{d-2}} \left[ \frac{8\Gamma(d/2)}{(d-1)\pi^{(d-2)/2}} \right]^{\frac{1}{d-2}} \left[1 - \frac{\bar{\alpha}(d-1)\pi^{(d-2)/2}}{2\Gamma(d/2)} + \frac{\bar{\beta}(d-1)^{2}\pi^{(d-2)}}{4\Gamma(d/2)^{2}}\right]^{\frac{1}{d-2}}.
\label{shadow_mass}
\end{equation}

Table~\ref{tab_shadow} presents the shadow radius results for the GUP linear and quadratic parameters $\bar{\alpha}$ and $\bar{\beta}$, used previously. Note that, in the uncorrected case, the shadow radius decreases drastically from $d=3$ ($5.196$) to $d=6$ ($1.216$). 
The linear GUP parameter, $\bar{\alpha}$, consistently reduces the shadow radius across all dimensions. For instance, at $d=3$ with $\bar{\beta}=0$ and $\bar{\alpha}$, varying from $0$ to $0.04$, the shadow radius decreases from $5.196$ to $4.780$. Conversely, the quadratic parameter $\bar{\beta}$ increases the shadow radius. For the same dimension $d=3$ with $\bar{\alpha}=0$ and $\bar{\beta}$ varying from $0$ to $0.03$, the shadow radius increases from  $5.196$ to $5.820$. This behavior is illustrated in Fig. \ref{Shadow}, where circles represent the shadow radius for each dimension, including the modifications resulting from the GUP correction. In the higher dimensions investigated, the quadratic component of the GUP correction is particularly interesting because it increases the shadow radius, enabling an analysis using EHT data to constrain the parameter $\bar{\beta}$.

\begin{table}[!ht]
\begin{center}
\caption{\footnotesize{Shadow radius $\mathcal{R}_{s}=R_{s}/M^{1/(d-2)}$ for different dimensions $d$ and GUP parameters $\bar{\alpha}$ and $\bar{\beta}$.}}
\label{tab_shadow}
\begin{footnotesize}
\begin{tabular}{|c|c|c|c|c|}
\hline
 $\mathcal{R}_{s}$ & $\bar{\alpha} = 0$,  $\bar{\beta} = 0 $ & $\bar{\alpha} = 0$, $\bar{\beta} = 0.03$ & $\bar{\alpha} = 0.04$, $  \bar{\beta} = 0  $ & $\bar{\alpha} = 0.04$, $\bar{\beta} = 0.03$ \\
\hline
$ d = 3$ & 5.19615 & 5.81969 & 4.78046 & 5.40400 \\
\hline
$ d = 4$ & 1.84264 & 2.37850 & 1.65991 & 2.23992 \\
\hline
$ d = 5$ & 1.36947 & 1.99800 & 1.19528 & 1.92339 \\
\hline
$ d = 6$ & 1.21629 & 1.86821 & 1.02610 & 1.82535 \\
\hline
\end{tabular}
\end{footnotesize}
\end{center}
\end{table}

\subsection{Quasinormal Modes and Shadow Correspondence}

A deep connection exists between the null geodesics and the quasinormal modes of black holes in the eikonal limit ($l \gg 1$). This correspondence, first established by Cardoso et al. \cite{Cardoso:2008bp}, states that the quasinormal frequencies are determined by the orbital angular frequency and the Lyapunov exponent associated with the unstable circular null orbit.
For our GUP-corrected Schwarzschild-Tangherlini black hole, the critical radius $r_c$, corresponding to the unstable circular null orbit, and the associated critical impact parameter $b_c$ are shown in equations \eqref{photon_sphere} and \eqref{critical_impact}, respectively. The angular velocity of the unstable circular null geodesic can be obtained by using the following relation:

\begin{equation}
\Omega_c = \dfrac{\dot{\phi}}{\dot{t}} = \frac{b_{c} B(r_c)}{r_{c}^{2}} = \frac{1}{b_c}.
\label{angular_velocity}
\end{equation}

The Lyapunov exponent, which measures the instability timescale, is given by:

\begin{equation}
\lambda_c = \sqrt{\frac{B(r_c)}{2r_{c}^{2}} \left( 2B(r_{c})-\left.\frac{d^2 B(r)}{dr^2}\right|_{r=r_c} \right) } = \Omega_c \sqrt{d-2}.
\label{lyapunov}
\end{equation}
Thus, the specific Schwarzschild-Tangherlini metric with GUP can alter the absolute scale of the quasinormal modes and the shadow, but not the ratio $\lambda_c/\Omega_{c}$ for this specific case of the metric.
According to the eikonal correspondence \cite{Cardoso:2008bp}, the QNM frequencies for $l \gg 1$ are:

\begin{equation}
\omega_{\text{QNM}} = \Omega_c l - i \left(n + \frac{1}{2}\right) |\lambda_c|, \qquad n = 0, 1, 2, \dots
\label{qnm_eikonal}
\end{equation}

\begin{figure}[!htb]
 \centering
 \subfigure[]{\includegraphics[scale=0.32]{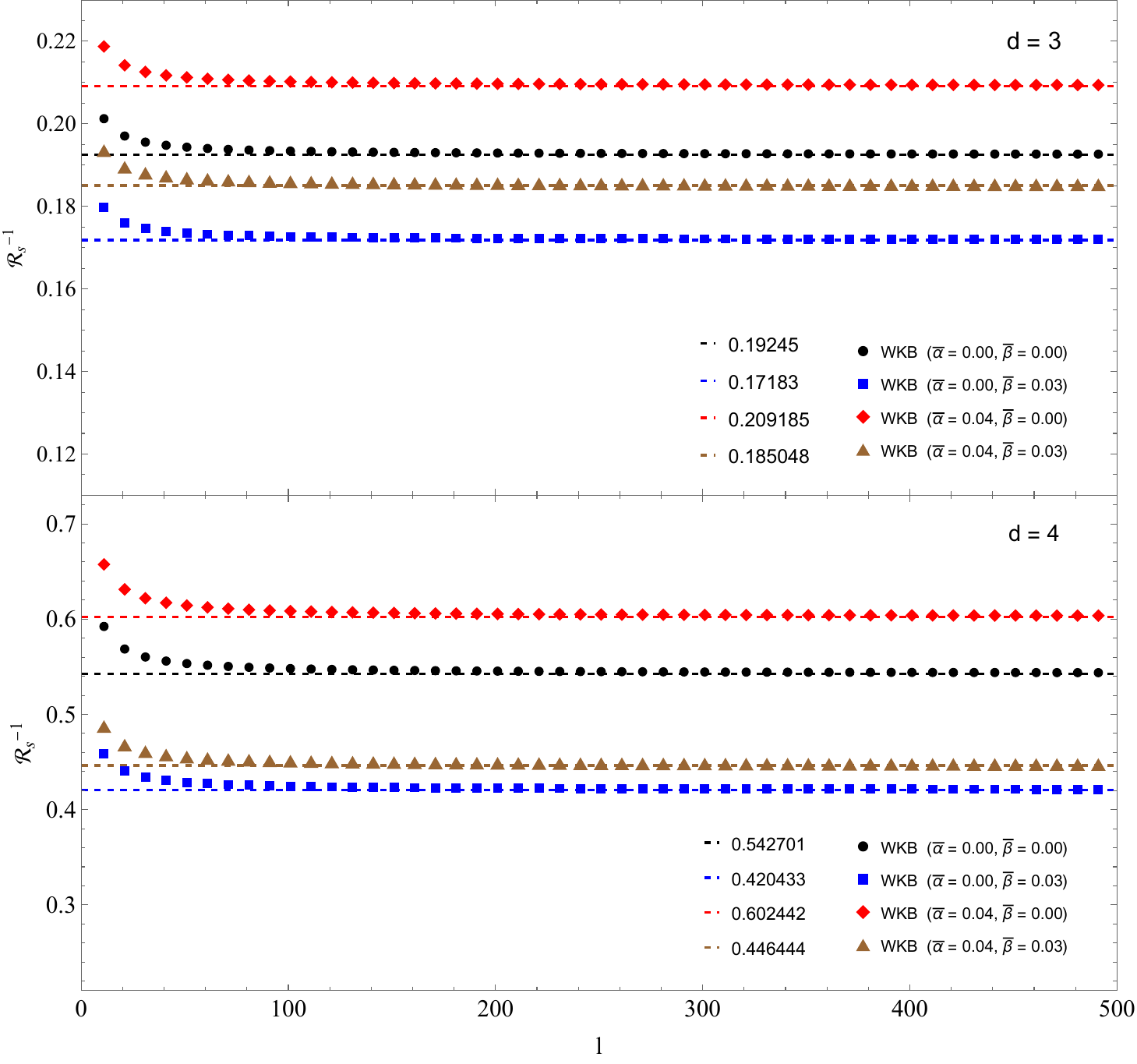}\label{Rs_WKBd3d4}}
 \qquad
 \subfigure[]{\includegraphics[scale=0.32]{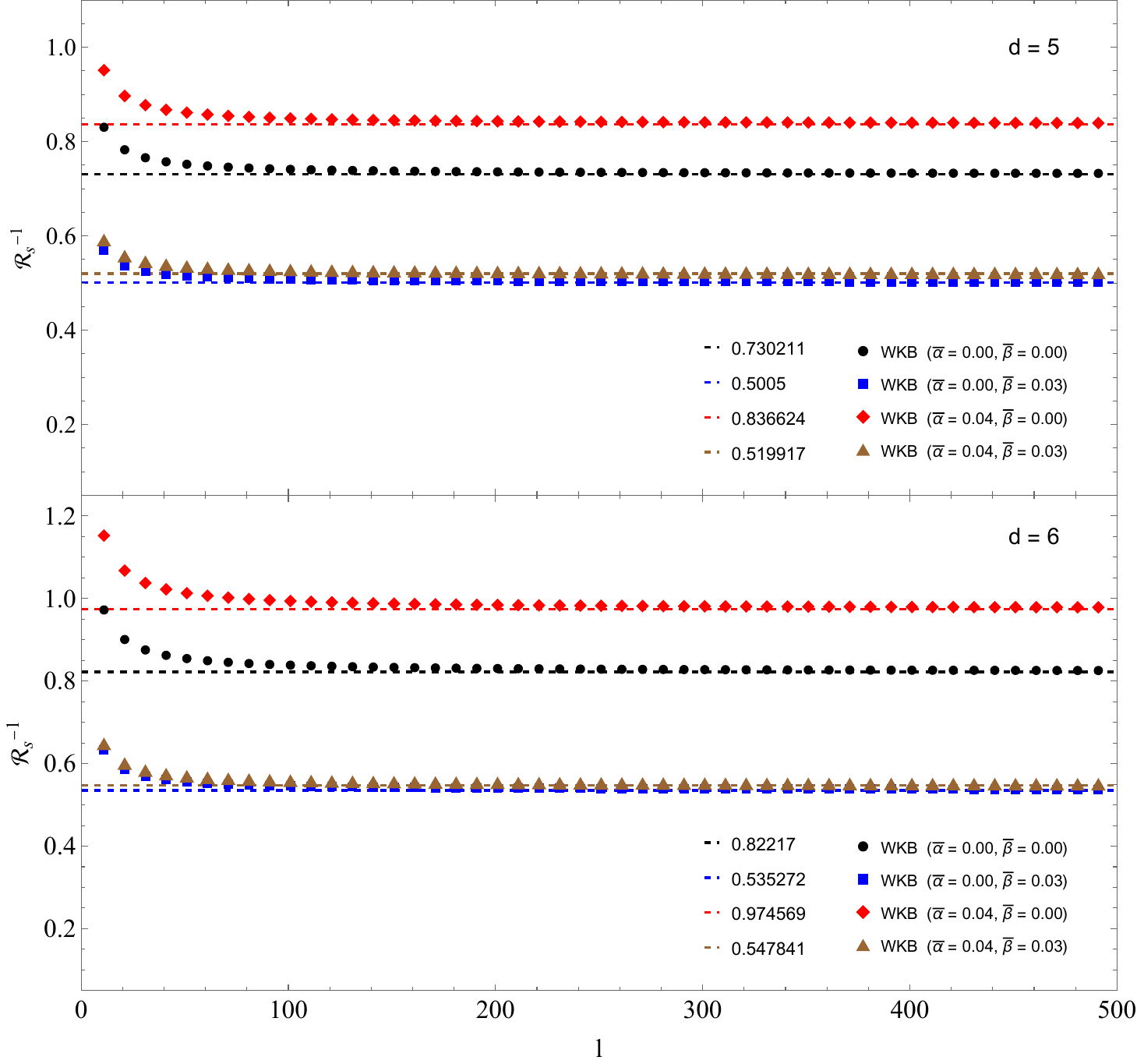}\label{Rs_WKBd5d6}}
 \caption{\footnotesize{The two panels show the behavior of the shadow radius for the dimensions under consideration, as well as the effects of the linear and quadratic GUP corrections. The dashed lines represent the shadow radius values obtained via equation \eqref{shadow_mass} compared with the results obtained using the WKB approximation. We assume the overtone $n = 0$ for this analysis.}}
  \label{Rs_WKB}
\end{figure}

Substituting \eqref{angular_velocity} and \eqref{lyapunov}, we obtain:

\begin{equation}
\omega_{\text{QNM}} = \frac{l}{b_c} - i \left(n + \frac{1}{2}\right) \frac{1}{b_c} \sqrt{d-2}.
\label{qnm_final}
\end{equation}
As we have seen, the shadow radius is related to the critical impact parameter by $R_{sh} = b_c$. Therefore,
\begin{equation}
\operatorname{Re}(\omega_{\text{QNM}}) = \lim_{l\rightarrow \infty} {\frac{l}{R_{s}}}, \qquad \operatorname{Im}(\omega_{\text{QNM}}) = -\left(n + \frac{1}{2}\right) \frac{1}{R_{s}} \sqrt{d-2}.
\label{qnm_shadow}
\end{equation}
Thus, in the eikonal limit, measuring the real part of the QNM frequency is directly related to the shadow radius:

\begin{equation}
R_{s} =  \lim_{l\rightarrow \infty} {\frac{l}{\operatorname{Re}(\omega_{\text{QNM}})}}.
\label{shadow_from_qnm}
\end{equation}
 In Figure~\ref{Rs_WKB}, we compare the result for the shadow radius with the real part of the quasinormal modes obtained using the WKB approximation. Note that for different values of dimension $d=3, 4, 5$ and $6$, the results approach each other in the limit of large $l\rightarrow \infty$.

\subsection{Constraints from EHT observations}

The Event Horizon Telescope (EHT) collaboration has published the first images of the supermassive black holes M87* \cite{akiyama2019first, eventhorizon2019first} and Sgr A* \cite{akiyama2022first}. These observations provide constraints on the angular size of the shadow, which can be used to test deviations from predictions of general relativity and to constrain parameters of modified gravitational models.
According to~\cite{EventHorizonTelescope:2022xqj}, the inferred constraints for the shadow size of Sgr A*, based on estimates from the Keck and VLTI instruments, are given at the $1\sigma$ level by:
\begin{eqnarray}
4.5\lesssim R_{s}/M_{\rm EHT}\lesssim5.5,\quad\text{and}\quad4.3\lesssim R_{s}/M_{\rm EHT}\lesssim 5.3.
\label{EHT}
\end{eqnarray}
where $M_{\rm EHT}$ denotes the gravitational radius associated with the mass of Sgr A* inferred from observations.
These measurements have been used to constrain deviations from general relativity predictions arising from extensions to the gravitational framework. Lemos et al.~\cite{Lemos:2024wwi} used the shadow of Sgr A* to investigate possible signatures of extra dimensions within the Randall-Sundrum scenario, obtaining constraints on the characteristic curvature scale of the extra dimensional spacetime. Previous studies using the shadow of M87* were conducted by Vagnozzi and Visinelli~\cite{Vagnozzi:2019apd} to constrain extra dimensional effects within the Randall-Sundrum framework.

In our work, we therefore define the associated characteristic gravitational length scale as $L_{M}\equiv M^{1/(d-2)}$. To compare the theoretical predictions with the EHT constraints, we adopt the phenomenological matching $M^{1/(d-2)}=M_{EHT}$. Under this identification, the observational constraints in \eqref{EHT} can be expressed as:
\begin{eqnarray}
4.5\lesssim R_{s}/M^{1/(d-2)}\lesssim 5.5,\quad\text{and}\quad4.3\lesssim R_{s}/M^{1/(d-2)}\lesssim5.3.
\label{EHT_para_Md}
\end{eqnarray}

\begin{figure}[htbp]
 \centering
\includegraphics[width=.45\textwidth]{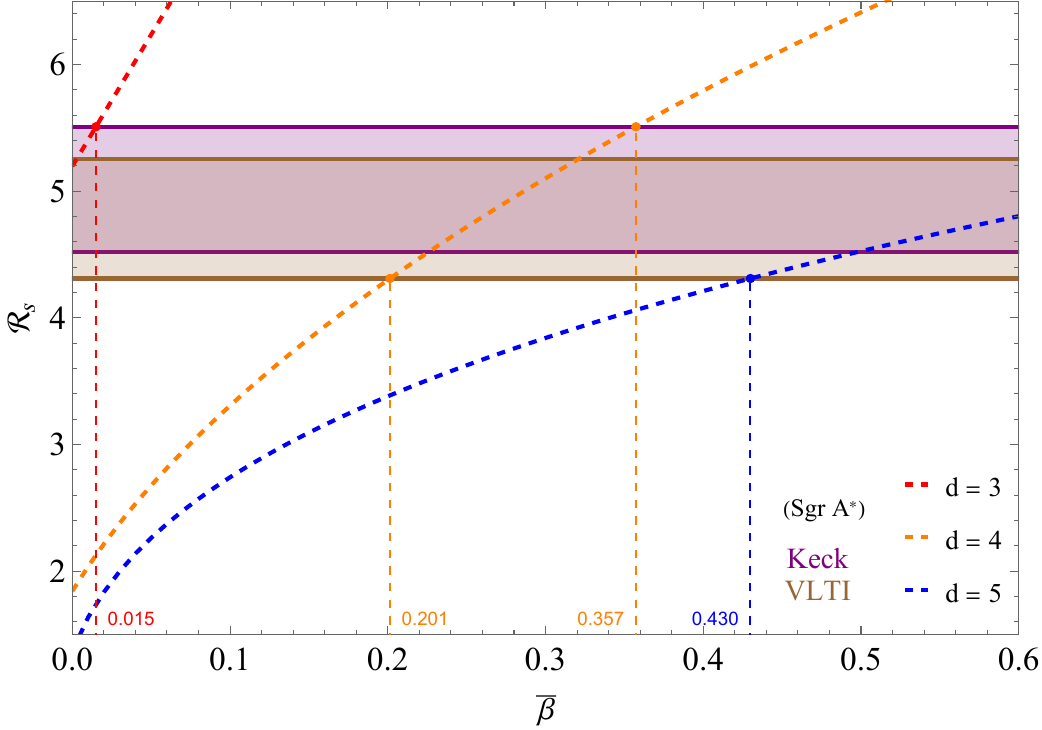}
  \caption{\footnotesize{Model shadow radius $\mathcal{R}_s=R_{s}/M^{1/(d-2)}$ compared with EHT data. For a fixed value of the linear GUP contribution $\bar{\alpha}=0$, there is a range of values of the quadratic term $\bar{\beta}$ that are allowed to be consistent with the data for $d=3,4$ and $5$.}}
 \label{plotRsSgrA}
\end{figure}

For a given set of parameters $\bar{\beta}$ and $\bar{\alpha}$ considered in Table~\ref{tab_shadow}, only the $d=3$ case lies within the observational ranges. For the calculated shadow radius to lie within the ranges allowed for consistency with EHT data, a sufficiently large quadratic GUP contribution is required, as shown in Figure~\ref{plotRsSgrA}.  The figure illustrates the behavior of the shadow radius for $d=3, 4$ and $5$. Assuming that the linear part of the GUP is fixed at $\bar{\alpha}=0$ while varying $\bar{\beta}$, we obtain the following constraints: for $d=4$ using Keck data, we find $0.226 \lesssim \bar{\beta} \lesssim 0.357$, while for VLTI data, we find $0.201 \lesssim \bar{\beta} \lesssim 0.320$. For $d=5$, the curve enters the allowed region when $\bar{\beta} \gtrsim 0.43$.

\begin{figure}[!htb]
 \centering
 \subfigure[]{\includegraphics[scale=0.4]{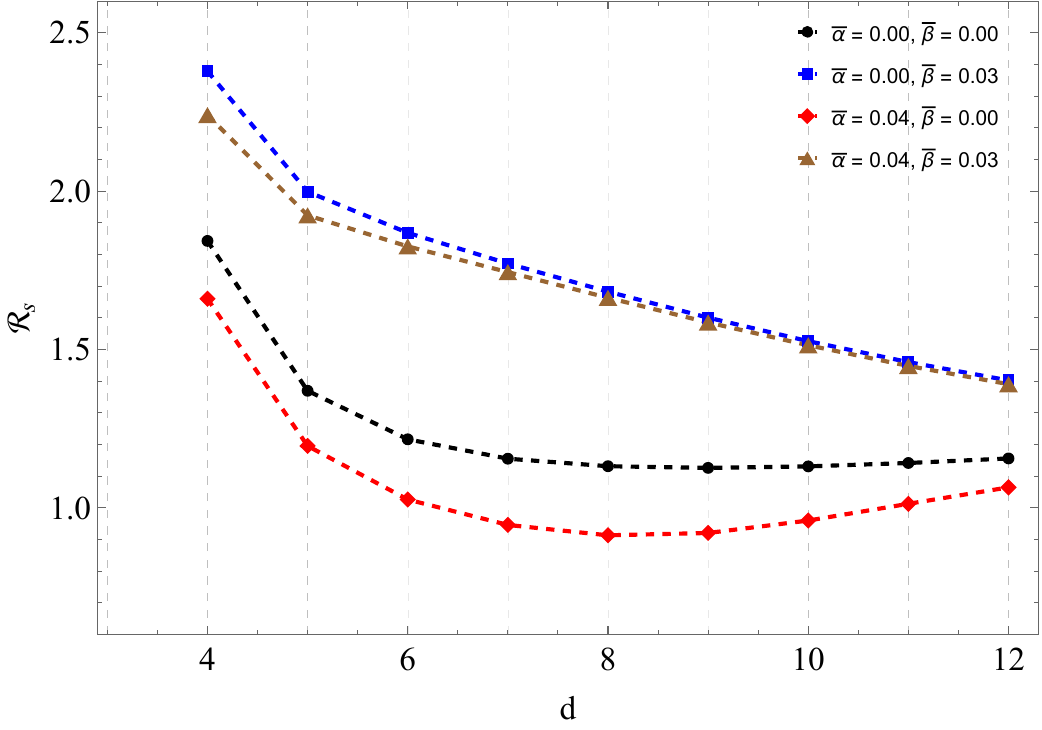}\label{Rsd4ad12}}
 \quad
 \subfigure[]{\includegraphics[scale=0.4]{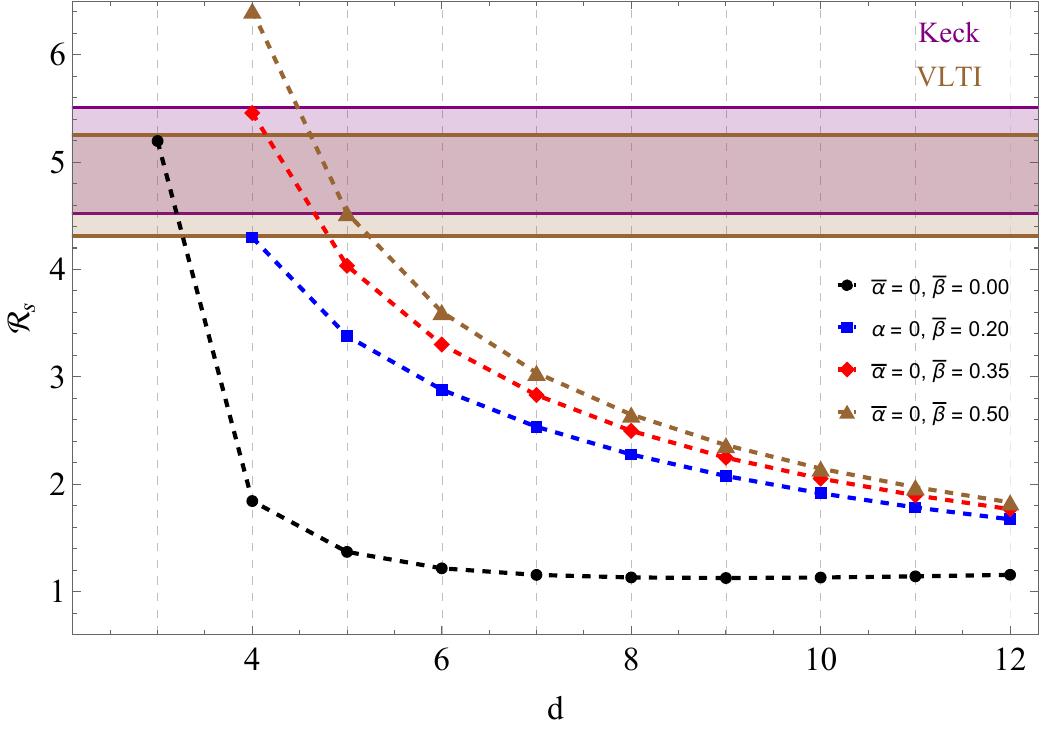}\label{Rsd4ad12Keck_VLTI}}
 \caption{\footnotesize{Dimensionless shadow radius $\mathcal{R}_s=R_{s}/M^{1/(d-2)}$ versus dimension $d$. In (a) shows the evolution with GUP parameters, showing a minimum at $d=8$ para $\bar{\alpha}=0.04$ and $\bar{\beta}=0$. In panel (b), we present values for $\bar{\beta}$ that adjust the shadow radius to the data region for $d=4$ and $d=5$.}}
  \label{Rs_d}
\end{figure}

Figure \ref{Rs_d} shows the behavior of the shadow radius for various values of $d$ and the influence of the GUP parameters. Note that the left panel presents a comparison using small values of $\bar{\alpha}$ and $\bar{\beta}$, the red curve shows the contribution solely from the linear term. We observe that the shadow radius decreases monotonically up to $d=8$, where it reaches a minimum, and then increases slightly for higher values of $d$, this behavior also appears, albeit more smoothly, in the uncorrected case (black). Conversely, the quadratic contribution raises the $\mathcal{R}_s$ values as seen in the blue and brown results but without producing a minimum within the analyzed range, instead reducing $\mathcal{R}_s$ as $d$ increases up to $d=12$. The presence of $\bar{\alpha}$ consistently reduces the shadow radius across all dimensions, reinforcing the idea that the linear correction acts as an effective reduction in horizon size.
The right panel \ref{Rsd4ad12Keck_VLTI} illustrates how values of $\bar{\beta}$ outside the range considered previously would be required to make higher-dimensional models compatible with the observational regions inferred from the EHT data.
For theoretical predictions in dimensions $d>3$ to fall within the observational regions, the contribution of the quadratic term $\bar{\beta}$ must lie beyond the parameter range adopted in our analysis thus far. For $d=4$, the minimum value of the quadratic GUP parameter needed to reach the observational regions is approximately $0.2\lesssim \bar{\beta} \lesssim 0.357$, depending on the observational range considered. 

\begin{figure}[!htb]
 \centering
 \subfigure[]{\includegraphics[scale=0.4]{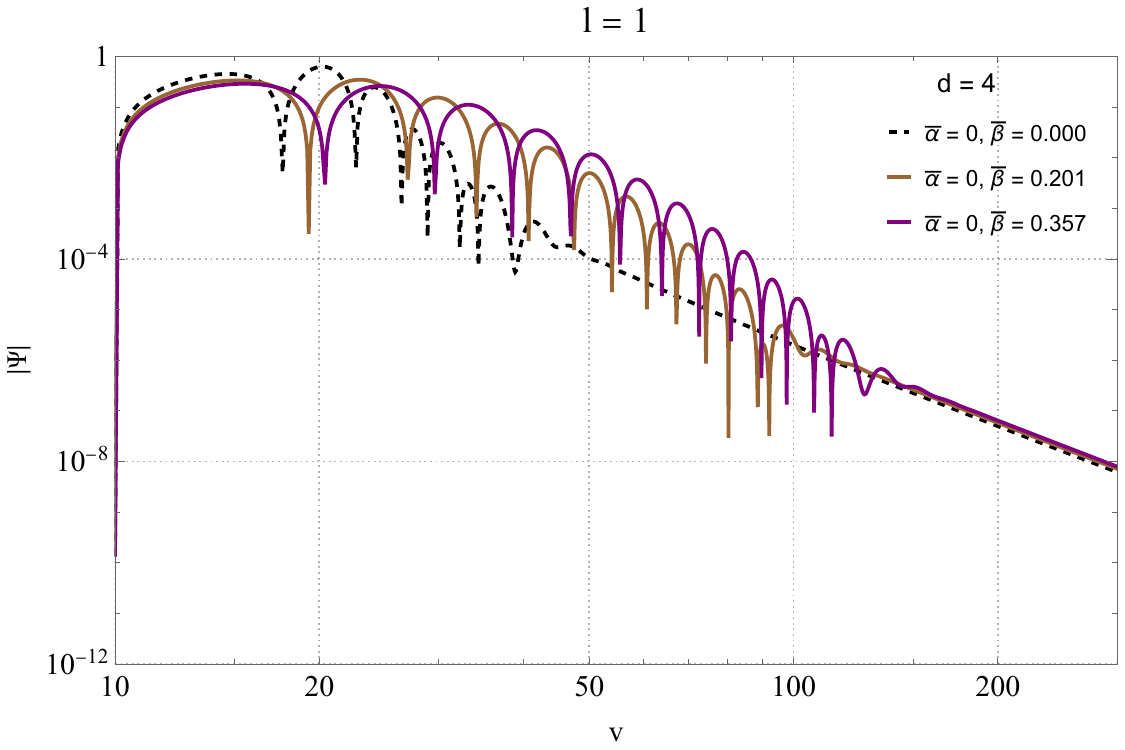}\label{MDFd4l1}}
 \quad
 \subfigure[]{\includegraphics[scale=0.4]{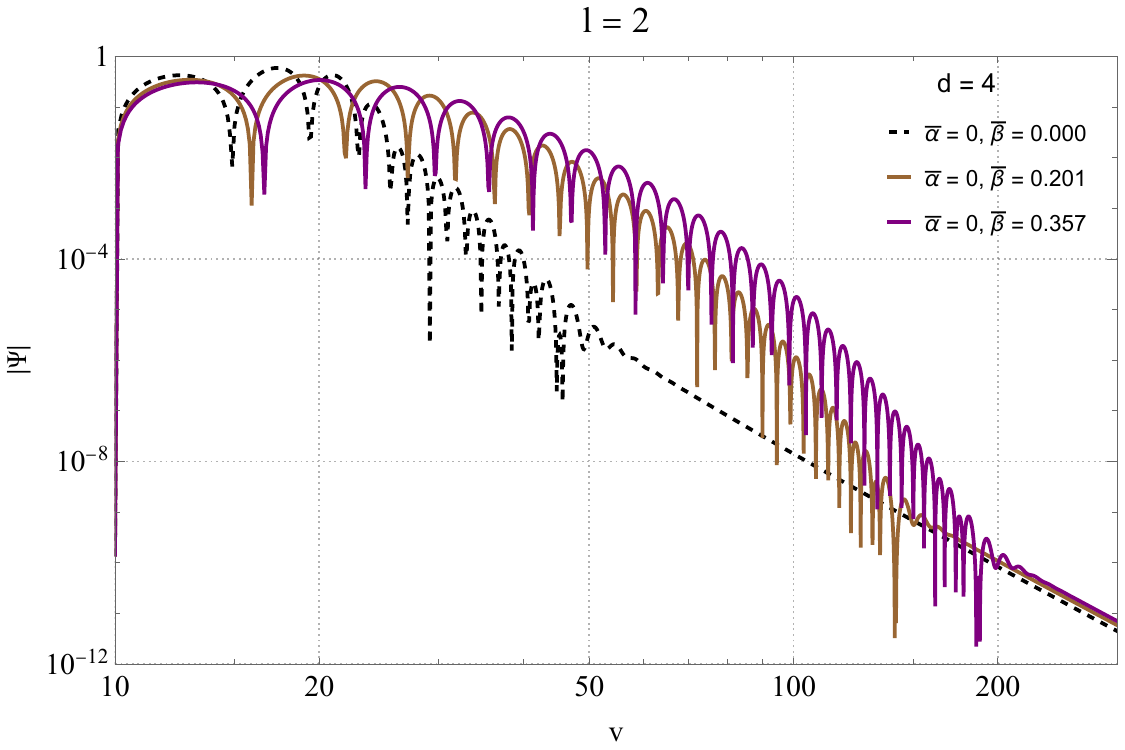}\label{MDFd4l2}}
 \caption{\footnotesize{Time-domain evolution for dimension $d=4$, at $l=1$ and $l=2$. The dashed black curves represent the uncorrected case Schwarzschild-Tangherlini, while the other two colors represent the limits obtained for $\bar{\beta}$ based on constraints from Keck (purple) and VLTI (brown) data.}}
\label{MDFd4}
\end{figure}

Relating the EHT constraints to the time-domain evolution discussed in Section \ref{s2}, Fig. \ref{MDFd4} displays the damped oscillations for $d=4$ corresponding to multipoles $l=1$ and $2$, using the constrained $\bar{\beta}$ values derived from the shadow analysis. 
The figure shows that, for $d=4$ without correction (dashed line), the damping is extremely rapid, in contrast, with restricted $\bar{\beta}$ values, the ringdown is longer and more structured, potentially producing appreciable modifications to the corresponding ringdown waveform. This behavior is consistent with the magnitude of the imaginary part of the quasinormal frequency shown in Table \ref{tab_WKBd4}.

Thus, the results suggest that the GUP corrections considered here can produce appreciable modifications to the quasinormal mode spectrum for $d=4$. Consequently, higher-dimensional scenarios would require increasingly large values of $\bar{\beta}$ within the adopted phenomenological normalization.

\begin{table}[!ht]
\begin{center}
\caption{\footnotesize{Quasinormal frequencies $\bar{\alpha} = 0$.}}
\label{tab_WKBd4}
\begin{footnotesize}
\begin{tabular}{|c|c|c|c|c|}
\hline
\multicolumn{5}{|c|}{$l = 1$}\\
\hline
$d=4$ & $  \hat{\omega}_{0} $ & $\hat{\omega}_{1}$ & $  \hat{\omega}_{2} $ & $\hat{\omega}_{3}$ \\
\hline
$ \bar{\beta} = 0$          & 1.101080 - 0.396434i & 0.928592 - 1.266000i & 0.691540 - 2.336840i & 0.512087 - 3.71482i \\
\hline
$ \beta = 0.201$ & 0.471066 - 0.169603i & 0.397272 - 0.541622i & 0.295856 - 0.999754i & 0.219082 - 1.58928i \\
\hline
$ \bar{\beta} = 0.357$ & 0.368508 - 0.132678 i & 0.31078 - 0.423703i & 0.231444 - 0.782093i & 0.171385 - 1.24327i \\
\hline
\hline
\multicolumn{5}{|c|}{$l = 2$}\\
\hline
$d=4$ & $  \hat{\omega}_{0} $ & $\hat{\omega}_{1}$ & $  \hat{\omega}_{2} $ & $\hat{\omega}_{3}$ \\
\hline
$ \bar{\beta} = 0$          & 1.639500 - 0.388248i & 1.511410 - 1.199770i & 1.287710 - 2.114670i & 1.022450 - 3.18851i \\
\hline
$ \bar{\beta} = 0.201$ & 0.701415 - 0.166101i & 0.646614 - 0.513288i & 0.550912 - 0.904705i & 0.437429 - 1.36412i \\
\hline
$ \bar{\beta} = 0.357$ & 0.548707 - 0.129938i & 0.505837 - 0.401538i & 0.430971 - 0.707738i & 0.342194 - 1.06713i \\
\hline
\end{tabular}
\end{footnotesize}
\end{center}
\end{table}
\section{Conclusions}
\label{conc}
In this work, we investigate the quasinormal spectrum and time-domain response of a Schwarzschild-Tangherlini black hole in the presence of Generalized Uncertainty Principle (GUP) corrections.
The quasinormal frequencies were obtained using the sixth-order WKB approximation. The results show a systematic dependence of the quasinormal modes on spacetime dimensionality, in the absence of GUP corrections, both the oscillation frequency $\mathrm{Re}(\hat{\omega})$ and the damping rate $|\mathrm{Im}(\hat{\omega})|$ increase with the number of dimensions. The same qualitative behavior is observed for both angular sectors, with higher values of $l$ leading to higher oscillation frequencies.

We have also investigated the influence of the linear and quadratic GUP parameters on the quasinormal spectrum. The linear correction controlled by $\bar{\alpha}$ shifts the quasinormal frequencies towards higher values for both the real part and the magnitude of the imaginary part, that is, the linear part of the correction is associated with faster oscillations and a shorter damping timescale. In contrast, the quadratic correction $\bar{\beta}$ produces within the parameter range considered a reduction in both the real part and the magnitude of the imaginary part of the quasinormal frequencies, thus the corresponding perturbations oscillate more slowly and decay over a longer characteristic timescale. When both corrections are included simultaneously, the resulting spectrum reflects a competition between the linear and quadratic contributions. This provides a characteristic theoretical signature of the two GUP contributions.

To complement the spectral analysis, we investigated the time-domain response of the scalar perturbations. The numerical evolution confirms that the GUP-induced modifications to the quasinormal spectrum are reflected in the oscillation and damping behavior of the waveform.
Importantly, the visibility of these modifications in the time-domain depends on the excitation coefficients of the quasinormal modes, which are governed by the initial perturbation profile. This observation is relevant to the dimensional dependence of the time-domain signal. As the number of dimensions increases, the faster the decay of the waveform amplitude, however, this reduction in the number of observable decay cycles should not be interpreted as evidence that the underlying GUP corrections become intrinsically weaker. Instead, it reflects the combined effect of the modified quasinormal spectrum, stronger dimensional damping, and the excitation properties of the initial perturbation.

Furthermore, we establish a correspondence between the quasinormal spectrum and the shadow of the GUP-corrected Schwarzschild-Tangherlini black hole in the eikonal regime. In the limit $l \gg 1$, the shadow radius and the real part of the quasinormal frequency are directly related through the angular frequency of the unstable circular null orbit. Thus, the comparison between the quasinormal frequencies derived via the WKB method and the shadow radius shows convergence as $l$ increases, providing an independent consistency check for the quasinormal analysis.
An important aspect of this work is the comparison with observations of the Sagittarius A* shadow obtained by the Event Horizon Telescope (EHT) collaboration. Using the observational ranges for the dimensionless shadow radius, we derive model dependent constraints on the quadratic GUP parameter by setting the linear contribution to $\bar{\alpha}=0$. For $d=3$, the allowed values of $\bar{\beta}$ fall within the range of a few tenths.
In the parameterization considered, the value of $\bar{\beta}$ required for the theoretical shadow to fall within the observational ranges increases with the dimensionality of spacetime.
For example, in the case $d=4$, the analysis yields $0.226\lesssim\bar{\beta}\lesssim 0.357$ for the Keck interval and $0.201\lesssim\bar{\beta}\lesssim 0.320$ for the VLTI interval, while for $d=5$, the shadow falls within the allowed region only for values of $\bar{\beta} \gtrsim 0.43$. 
Following the phenomenological mass normalization adopted here, the comparison with the shadow of Sgr A* places a constraint on the magnitude of the quadratic GUP parameter, with the required values of $\bar{\beta}$ significantly depending on the dimensionality of spacetime.

{\acknowledgments We thank CNPq and CAPES for partial financial support. MAA, FAB and EP acknowledge support from CNPq (grant nos. $301683/2025-5$, $309092/2022-1$ and $304290/2020-3$). JAVC and ARQ thank the Paraíba State Research Support Foundation (FAPESQ) for financial support. ARQ acknowledges the financial support by CNPq under process number $306884/2026-7$.}

\bibliographystyle{plainurl}

\begin{thebibliography}{10}

\bibitem{abac2025gw250114}
A.~G. Abac et~al.
\newblock {GW250114: Testing Hawking{\textquoteright}s Area Law and the Kerr
  Nature of Black Holes}.
\newblock {\em Phys. Rev. Lett.}, 135(11):111403, 2025.
\newblock \href {http://arxiv.org/abs/2509.08054} {\path{arXiv:2509.08054}},
  \href {https://doi.org/10.1103/kw5g-d732} {\path{doi:10.1103/kw5g-d732}}.

\bibitem{abac2026black}
A.~G. Abac et~al.
\newblock {Black Hole Spectroscopy and Tests of General Relativity with
  GW250114}.
\newblock {\em Phys. Rev. Lett.}, 136(4):041403, 2026.
\newblock \href {http://arxiv.org/abs/2509.08099} {\path{arXiv:2509.08099}},
  \href {https://doi.org/10.1103/6c61-fm1n} {\path{doi:10.1103/6c61-fm1n}}.

\bibitem{abbott2016tests}
B.~P. Abbott et~al.
\newblock {Tests of general relativity with GW150914}.
\newblock {\em Phys. Rev. Lett.}, 116(22):221101, 2016.
\newblock [Erratum: Phys.Rev.Lett. 121, 129902 (2018)].
\newblock \href {http://arxiv.org/abs/1602.03841} {\path{arXiv:1602.03841}},
  \href {https://doi.org/10.1103/PhysRevLett.116.221101}
  {\path{doi:10.1103/PhysRevLett.116.221101}}.

\bibitem{Ahmed:2025boj}
Fazlay Ahmed, Heena Ali, Qiang Wu, Tao Zhu, and Sushant~G. Ghosh.
\newblock {Shadows of rotating non-commutative Kiselev black holes: constraints
  from EHT observations of M87* and Sgr A*}.
\newblock {\em Eur. Phys. J. C}, 85(7):795, 2025.
\newblock \href {https://doi.org/10.1140/epjc/s10052-025-14510-5}
  {\path{doi:10.1140/epjc/s10052-025-14510-5}}.

\bibitem{akiyama2019first}
Kazunori Akiyama et~al.
\newblock {First M87 Event Horizon Telescope Results. I. The Shadow of the
  Supermassive Black Hole}.
\newblock {\em Astrophys. J. Lett.}, 875:L1, 2019.
\newblock \href {http://arxiv.org/abs/1906.11238} {\path{arXiv:1906.11238}},
  \href {https://doi.org/10.3847/2041-8213/ab0ec7}
  {\path{doi:10.3847/2041-8213/ab0ec7}}.

\bibitem{eventhorizon2019first}
Kazunori Akiyama et~al.
\newblock {First M87 Event Horizon Telescope Results. IV. Imaging the Central
  Supermassive Black Hole}.
\newblock {\em Astrophys. J. Lett.}, 875(1):L4, 2019.
\newblock \href {http://arxiv.org/abs/1906.11241} {\path{arXiv:1906.11241}},
  \href {https://doi.org/10.3847/2041-8213/ab0e85}
  {\path{doi:10.3847/2041-8213/ab0e85}}.

\bibitem{akiyama2022first}
Kazunori Akiyama et~al.
\newblock {First Sagittarius A* Event Horizon Telescope Results. I. The Shadow
  of the Supermassive Black Hole in the Center of the Milky Way}.
\newblock {\em Astrophys. J. Lett.}, 930(2):L12, 2022.
\newblock \href {http://arxiv.org/abs/2311.08680} {\path{arXiv:2311.08680}},
  \href {https://doi.org/10.3847/2041-8213/ac6674}
  {\path{doi:10.3847/2041-8213/ac6674}}.

\bibitem{EventHorizonTelescope:2022xqj}
Kazunori Akiyama et~al.
\newblock {First Sagittarius A* Event Horizon Telescope Results. VI. Testing
  the Black Hole Metric}.
\newblock {\em Astrophys. J. Lett.}, 930(2):L17, 2022.
\newblock \href {http://arxiv.org/abs/2311.09484} {\path{arXiv:2311.09484}},
  \href {https://doi.org/10.3847/2041-8213/ac6756}
  {\path{doi:10.3847/2041-8213/ac6756}}.

\bibitem{Al-Badawi:2024cby}
Ahmad Al-Badawi, Sanjar Shaymatov, Sohan~Kumar Jha, and Anisur Rahaman.
\newblock {GUP corrected black holes with cloud of string}.
\newblock {\em Eur. Phys. J. C}, 84(7):722, 2024.
\newblock \href {http://arxiv.org/abs/2406.17501} {\path{arXiv:2406.17501}},
  \href {https://doi.org/10.1140/epjc/s10052-024-13059-z}
  {\path{doi:10.1140/epjc/s10052-024-13059-z}}.

\bibitem{Ali:2009zq}
Ahmed~Farag Ali, Saurya Das, and Elias~C. Vagenas.
\newblock {Discreteness of Space from the Generalized Uncertainty Principle}.
\newblock {\em Phys. Lett. B}, 678:497--499, 2009.
\newblock \href {http://arxiv.org/abs/0906.5396} {\path{arXiv:0906.5396}},
  \href {https://doi.org/10.1016/j.physletb.2009.06.061}
  {\path{doi:10.1016/j.physletb.2009.06.061}}.

\bibitem{Ali:2024ssf}
Heena Ali, Shafqat~Ul Islam, and Sushant~G. Ghosh.
\newblock {Shadows and parameter estimation of rotating quantum corrected black
  holes and constraints from EHT observation of M87* and Sgr A*}.
\newblock {\em JHEAp}, 47:100367, 2025.
\newblock \href {http://arxiv.org/abs/2410.09198} {\path{arXiv:2410.09198}},
  \href {https://doi.org/10.1016/j.jheap.2025.100367}
  {\path{doi:10.1016/j.jheap.2025.100367}}.

\bibitem{Anacleto:2020lel}
M.~A. Anacleto, F.~A. Brito, J.~A.~V. Campos, and E.~Passos.
\newblock {Quantum-corrected scattering and absorption of a Schwarzschild black
  hole with GUP}.
\newblock {\em Phys. Lett. B}, 810:135830, 2020.
\newblock \href {http://arxiv.org/abs/2003.13464} {\path{arXiv:2003.13464}},
  \href {https://doi.org/10.1016/j.physletb.2020.135830}
  {\path{doi:10.1016/j.physletb.2020.135830}}.

\bibitem{Anacleto:2023ntm}
M.~A. Anacleto, J.~A.~V. Campos, F.~A. Brito, E.~Maciel, and E.~Passos.
\newblock {Scattering and absorption by extra-dimensional black holes with
  GUP}.
\newblock {\em Nucl. Phys. B}, 1006:116617, 2024.
\newblock \href {http://arxiv.org/abs/2307.09536} {\path{arXiv:2307.09536}},
  \href {https://doi.org/10.1016/j.nuclphysb.2024.116617}
  {\path{doi:10.1016/j.nuclphysb.2024.116617}}.

\bibitem{Anacleto:2021qoe}
M.~A. Anacleto, J.~A.~V. Campos, F.~A. Brito, and E.~Passos.
\newblock {Quasinormal modes and shadow of a Schwarzschild black hole with
  GUP}.
\newblock {\em Annals Phys.}, 434:168662, 2021.
\newblock \href {http://arxiv.org/abs/2108.04998} {\path{arXiv:2108.04998}},
  \href {https://doi.org/10.1016/j.aop.2021.168662}
  {\path{doi:10.1016/j.aop.2021.168662}}.

\bibitem{Barman:2024hwd}
Himangshu Barman, Ahmad Al-Badawi, Sohan~Kumar Jha, and Anisur Rahaman.
\newblock {The quantum corrected Schwarzschild black hole with a
  linear-quadratic GUP: a comprehensive evaluation}.
\newblock {\em JCAP}, 05:019, 2024.
\newblock \href {http://arxiv.org/abs/2401.14833} {\path{arXiv:2401.14833}},
  \href {https://doi.org/10.1088/1475-7516/2024/05/019}
  {\path{doi:10.1088/1475-7516/2024/05/019}}.

\bibitem{Benda:2025tni}
Kristian Benda and Jerzy Matyjasek.
\newblock {Quasinormal modes of black holes: Efficient and highly accurate
  calculations with recurrence-based methods}.
\newblock {\em Phys. Rev. D}, 111(12):124010, 2025.
\newblock \href {http://arxiv.org/abs/2503.17325} {\path{arXiv:2503.17325}},
  \href {https://doi.org/10.1103/2bg3-t96k} {\path{doi:10.1103/2bg3-t96k}}.

\bibitem{Berti:2009kk}
Emanuele Berti, Vitor Cardoso, and Andrei~O. Starinets.
\newblock {Quasinormal modes of black holes and black branes}.
\newblock {\em Class. Quant. Grav.}, 26:163001, 2009.
\newblock \href {http://arxiv.org/abs/0905.2975} {\path{arXiv:0905.2975}},
  \href {https://doi.org/10.1088/0264-9381/26/16/163001}
  {\path{doi:10.1088/0264-9381/26/16/163001}}.

\bibitem{Campos:2021sff}
J.~A.~V. Campos, M.~A. Anacleto, F.~A. Brito, and E.~Passos.
\newblock {Quasinormal modes and shadow of noncommutative black hole}.
\newblock {\em Sci. Rep.}, 12(1):8516, 2022.
\newblock \href {http://arxiv.org/abs/2103.10659} {\path{arXiv:2103.10659}},
  \href {https://doi.org/10.1038/s41598-022-12343-w}
  {\path{doi:10.1038/s41598-022-12343-w}}.

\bibitem{Cardoso:2003vt}
Vitor Cardoso, Jose P.~S. Lemos, and Shijun Yoshida.
\newblock {Quasinormal modes of Schwarzschild black holes in four-dimensions
  and higher dimensions}.
\newblock {\em Phys. Rev. D}, 69:044004, 2004.
\newblock \href {http://arxiv.org/abs/gr-qc/0309112}
  {\path{arXiv:gr-qc/0309112}}, \href
  {https://doi.org/10.1103/PhysRevD.69.044004}
  {\path{doi:10.1103/PhysRevD.69.044004}}.

\bibitem{Cardoso:2008bp}
Vitor Cardoso, Alex~S. Miranda, Emanuele Berti, Helvi Witek, and Vilson~T.
  Zanchin.
\newblock {Geodesic stability, Lyapunov exponents and quasinormal modes}.
\newblock {\em Phys. Rev. D}, 79(6):064016, 2009.
\newblock \href {http://arxiv.org/abs/0812.1806} {\path{arXiv:0812.1806}},
  \href {https://doi.org/10.1103/PhysRevD.79.064016}
  {\path{doi:10.1103/PhysRevD.79.064016}}.

\bibitem{Chen:2023wkq}
H.~Chen, T.~Sathiyaraj, H.~Hassanabadi, Y.~Yang, Z.~W. Long, and F.~Q. Tu.
\newblock {Quasinormal modes of the EGUP-corrected Schwarzschild black hole}.
\newblock {\em Indian J. Phys.}, 97(14):4481--4489, 2023.
\newblock \href {https://doi.org/10.1007/s12648-023-02734-8}
  {\path{doi:10.1007/s12648-023-02734-8}}.

\bibitem{Feng:2015jlj}
Z.~W. Feng, H.~L. Li, X.~T. Zu, and S.~Z. Yang.
\newblock {Corrections to the thermodynamics of Schwarzschild-Tangherlini black
  hole and the generalized uncertainty principle}.
\newblock {\em Eur. Phys. J. C}, 76(4):212, 2016.
\newblock \href {http://arxiv.org/abs/1604.04702} {\path{arXiv:1604.04702}},
  \href {https://doi.org/10.1140/epjc/s10052-016-4057-1}
  {\path{doi:10.1140/epjc/s10052-016-4057-1}}.

\bibitem{Gulia:2025tvx}
Himanshi Gulia, J.~K. Singh, Farruh Atamurotov, and Sushant~G. Ghosh.
\newblock {Observational signatures of shadows in GUP-corrected Kerr black
  holes and constraints from EHT data}.
\newblock {\em Phys. Dark Univ.}, 48:101954, 2025.
\newblock \href {https://doi.org/10.1016/j.dark.2025.101954}
  {\path{doi:10.1016/j.dark.2025.101954}}.

\bibitem{Gundlach:1993tn}
Carsten Gundlach, Richard~H. Price, and Jorge Pullin.
\newblock {Late time behavior of stellar collapse and explosions: 2. Nonlinear
  evolution}.
\newblock {\em Phys. Rev. D}, 49:890--899, 1994.
\newblock \href {http://arxiv.org/abs/gr-qc/9307010}
  {\path{arXiv:gr-qc/9307010}}, \href {https://doi.org/10.1103/PhysRevD.49.890}
  {\path{doi:10.1103/PhysRevD.49.890}}.

\bibitem{Han:2025cal}
Hyewon Han and Bogeun Gwak.
\newblock {Correspondence between quasinormal modes and greybody factors in
  five-dimensional black holes}.
\newblock {\em Phys. Rev. D}, 113(6):064058, 2026.
\newblock \href {http://arxiv.org/abs/2508.12989} {\path{arXiv:2508.12989}},
  \href {https://doi.org/10.1103/n2ns-drkp} {\path{doi:10.1103/n2ns-drkp}}.

\bibitem{Karmakar:2022idu}
Ronit Karmakar, Dhruba~Jyoti Gogoi, and Umananda~Dev Goswami.
\newblock {Quasinormal modes and thermodynamic properties of GUP-corrected
  Schwarzschild black hole surrounded by quintessence}.
\newblock {\em Int. J. Mod. Phys. A}, 37(28n29):2250180, 2022.
\newblock \href {http://arxiv.org/abs/2206.09081} {\path{arXiv:2206.09081}},
  \href {https://doi.org/10.1142/S0217751X22501809}
  {\path{doi:10.1142/S0217751X22501809}}.

\bibitem{Kempf:1994su}
Achim Kempf, Gianpiero Mangano, and Robert~B. Mann.
\newblock {Hilbert space representation of the minimal length uncertainty
  relation}.
\newblock {\em Phys. Rev. D}, 52:1108--1118, 1995.
\newblock \href {http://arxiv.org/abs/hep-th/9412167}
  {\path{arXiv:hep-th/9412167}}, \href
  {https://doi.org/10.1103/PhysRevD.52.1108}
  {\path{doi:10.1103/PhysRevD.52.1108}}.

\bibitem{Koch:2025gaw}
Benjamin Koch, Gonzalo~J. Olmo, Ali Riahinia, {\'A}ngel Rinc{\'o}n, and Diego
  Rubiera-Garcia.
\newblock {Quasi-normal modes and shadows of scale-dependent regular black
  holes}.
\newblock {\em JCAP}, 03:048, 2026.
\newblock \href {http://arxiv.org/abs/2506.15944} {\path{arXiv:2506.15944}},
  \href {https://doi.org/10.1088/1475-7516/2026/03/048}
  {\path{doi:10.1088/1475-7516/2026/03/048}}.

\bibitem{Konoplya:2003ii}
R.~A. Konoplya.
\newblock {Quasinormal behavior of the d-dimensional Schwarzschild black hole
  and higher order WKB approach}.
\newblock {\em Phys. Rev. D}, 68:024018, 2003.
\newblock \href {http://arxiv.org/abs/gr-qc/0303052}
  {\path{arXiv:gr-qc/0303052}}, \href
  {https://doi.org/10.1103/PhysRevD.68.024018}
  {\path{doi:10.1103/PhysRevD.68.024018}}.

\bibitem{Konoplya:2024lch}
R.~A. Konoplya and O.~S. Stashko.
\newblock {Probing the effective quantum gravity via quasinormal modes and
  shadows of black holes}.
\newblock {\em Phys. Rev. D}, 111(10):104055, 2025.
\newblock \href {http://arxiv.org/abs/2408.02578} {\path{arXiv:2408.02578}},
  \href {https://doi.org/10.1103/PhysRevD.111.104055}
  {\path{doi:10.1103/PhysRevD.111.104055}}.

\bibitem{Konoplya:2011qq}
R.~A. Konoplya and A.~Zhidenko.
\newblock {Quasinormal modes of black holes: From astrophysics to string
  theory}.
\newblock {\em Rev. Mod. Phys.}, 83:793--836, 2011.
\newblock \href {http://arxiv.org/abs/1102.4014} {\path{arXiv:1102.4014}},
  \href {https://doi.org/10.1103/RevModPhys.83.793}
  {\path{doi:10.1103/RevModPhys.83.793}}.

\bibitem{Konoplya:2019hlu}
R.~A. Konoplya, A.~Zhidenko, and A.~F. Zinhailo.
\newblock {Higher order WKB formula for quasinormal modes and grey-body
  factors: recipes for quick and accurate calculations}.
\newblock {\em Class. Quant. Grav.}, 36:155002, 2019.
\newblock \href {http://arxiv.org/abs/1904.10333} {\path{arXiv:1904.10333}},
  \href {https://doi.org/10.1088/1361-6382/ab2e25}
  {\path{doi:10.1088/1361-6382/ab2e25}}.

\bibitem{Lambiase:2023hng}
Gaetano Lambiase, Reggie~C. Pantig, Dhruba~Jyoti Gogoi, and Ali {\"O}vg{\"u}n.
\newblock {Investigating the connection between generalized uncertainty
  principle and asymptotically safe gravity in black hole signatures through
  shadow and quasinormal modes}.
\newblock {\em Eur. Phys. J. C}, 83(7):679, 2023.
\newblock \href {http://arxiv.org/abs/2304.00183} {\path{arXiv:2304.00183}},
  \href {https://doi.org/10.1140/epjc/s10052-023-11848-6}
  {\path{doi:10.1140/epjc/s10052-023-11848-6}}.

\bibitem{Lemos:2024wwi}
A.~S. Lemos, J.~A.~V. Campos, and F.~A. Brito.
\newblock {Hunting for extra dimensions in black hole shadows}.
\newblock {\em Phys. Rev. D}, 110(6):064079, 2024.
\newblock \href {http://arxiv.org/abs/2407.04609} {\path{arXiv:2407.04609}},
  \href {https://doi.org/10.1103/PhysRevD.110.064079}
  {\path{doi:10.1103/PhysRevD.110.064079}}.

\bibitem{Liu:2020ola}
Cheng Liu, Tao Zhu, Qiang Wu, Kimet Jusufi, Mubasher Jamil, Mustapha
  Azreg-A{\"\i}nou, and Anzhong Wang.
\newblock {Shadow and quasinormal modes of a rotating loop quantum black hole}.
\newblock {\em Phys. Rev. D}, 101(8):084001, 2020.
\newblock [Erratum: Phys.Rev.D 103, 089902 (2021)].
\newblock \href {http://arxiv.org/abs/2003.00477} {\path{arXiv:2003.00477}},
  \href {https://doi.org/10.1103/PhysRevD.101.084001}
  {\path{doi:10.1103/PhysRevD.101.084001}}.

\bibitem{Matyjasek:2021xfg}
Jerzy Matyjasek.
\newblock {Accurate quasinormal modes of the five-dimensional
  Schwarzschild-Tangherlini black holes}.
\newblock {\em Phys. Rev. D}, 104(8):084066, 2021.
\newblock \href {http://arxiv.org/abs/2107.04815} {\path{arXiv:2107.04815}},
  \href {https://doi.org/10.1103/PhysRevD.104.084066}
  {\path{doi:10.1103/PhysRevD.104.084066}}.

\bibitem{Matyjasek:2026yiu}
Jerzy Matyjasek, Roman~A. Konoplya, and Alexander Zhidenko.
\newblock {An Efficient Higher-Order WKB Code for Quasinormal Modes and
  Greybody Factors}.
\newblock {\em Int. J. Grav. Theor. Phys.}, 2(1):5, 2026.
\newblock \href {http://arxiv.org/abs/2603.12466} {\path{arXiv:2603.12466}},
  \href {https://doi.org/10.53941/ijgtp.2026.100005}
  {\path{doi:10.53941/ijgtp.2026.100005}}.

\bibitem{Pedrotti:2024znu}
Davide Pedrotti and Sunny Vagnozzi.
\newblock {Quasinormal modes-shadow correspondence for rotating regular black
  holes}.
\newblock {\em Phys. Rev. D}, 110(8):084075, 2024.
\newblock \href {http://arxiv.org/abs/2404.07589} {\path{arXiv:2404.07589}},
  \href {https://doi.org/10.1103/PhysRevD.110.084075}
  {\path{doi:10.1103/PhysRevD.110.084075}}.

\bibitem{Raza:2025ohk}
Muhammad~Ali Raza, M.~Zubair, Farruh Atamurotov, and Ahmadjon Abdujabbarov.
\newblock {Influence of quantum correction on Kerr black hole in effective loop
  quantum gravity via shadows and EHT results}.
\newblock {\em Eur. Phys. J. C}, 85(9):973, 2025.
\newblock \href {http://arxiv.org/abs/2501.01308} {\path{arXiv:2501.01308}},
  \href {https://doi.org/10.1140/epjc/s10052-025-14666-0}
  {\path{doi:10.1140/epjc/s10052-025-14666-0}}.

\bibitem{Sadhu:2018zyh}
Amruta Sadhu and Vardarajan Suneeta.
\newblock {Schwarzschild-Tangherlini quasinormal modes at large $D$ revisited}.
\newblock 6 2018.
\newblock \href {http://arxiv.org/abs/1806.04888} {\path{arXiv:1806.04888}}.

\bibitem{Singh:2017vfr}
Balendra~Pratap Singh and Sushant~G. Ghosh.
\newblock {Shadow of Schwarzschild{\textendash}Tangherlini black holes}.
\newblock {\em Annals Phys.}, 395:127--137, 2018.
\newblock \href {http://arxiv.org/abs/1707.07125} {\path{arXiv:1707.07125}},
  \href {https://doi.org/10.1016/j.aop.2018.05.010}
  {\path{doi:10.1016/j.aop.2018.05.010}}.

\bibitem{Tangherlini:1963bw}
F.~R. Tangherlini.
\newblock {Schwarzschild field in n dimensions and the dimensionality of space
  problem}.
\newblock {\em Nuovo Cim.}, 27:636--651, 1963.
\newblock \href {https://doi.org/10.1007/BF02784569}
  {\path{doi:10.1007/BF02784569}}.

\bibitem{Tsukamoto:2014dta}
Naoki Tsukamoto, Takao Kitamura, Koki Nakajima, and Hideki Asada.
\newblock {Gravitational lensing in Tangherlini spacetime in the weak
  gravitational field and the strong gravitational field}.
\newblock {\em Phys. Rev. D}, 90(6):064043, 2014.
\newblock \href {http://arxiv.org/abs/1402.6823} {\path{arXiv:1402.6823}},
  \href {https://doi.org/10.1103/PhysRevD.90.064043}
  {\path{doi:10.1103/PhysRevD.90.064043}}.

\bibitem{Vagnozzi:2022moj}
Sunny Vagnozzi et~al.
\newblock {Horizon-scale tests of gravity theories and fundamental physics from
  the Event Horizon Telescope image of Sagittarius A}.
\newblock {\em Class. Quant. Grav.}, 40(16):165007, 2023.
\newblock \href {http://arxiv.org/abs/2205.07787} {\path{arXiv:2205.07787}},
  \href {https://doi.org/10.1088/1361-6382/acd97b}
  {\path{doi:10.1088/1361-6382/acd97b}}.

\bibitem{Vagnozzi:2019apd}
Sunny Vagnozzi and Luca Visinelli.
\newblock {Hunting for extra dimensions in the shadow of M87*}.
\newblock {\em Phys. Rev. D}, 100(2):024020, 2019.
\newblock \href {http://arxiv.org/abs/1905.12421} {\path{arXiv:1905.12421}},
  \href {https://doi.org/10.1103/PhysRevD.100.024020}
  {\path{doi:10.1103/PhysRevD.100.024020}}.

\bibitem{Vazquez:2003zm}
Samuel~E. Vazquez and Ernesto~P. Esteban.
\newblock {Strong field gravitational lensing by a Kerr black hole}.
\newblock {\em Nuovo Cim. B}, 119:489--519, 2004.
\newblock \href {http://arxiv.org/abs/gr-qc/0308023}
  {\path{arXiv:gr-qc/0308023}}, \href
  {https://doi.org/10.1393/ncb/i2004-10121-y}
  {\path{doi:10.1393/ncb/i2004-10121-y}}.

\bibitem{Vieira:2023ylz}
H.~S. Vieira, Kyriakos Destounis, and Kostas~D. Kokkotas.
\newblock {Analog Schwarzschild black holes of Bose-Einstein condensates in a
  cavity: Quasinormal modes and quasibound states}.
\newblock {\em Phys. Rev. D}, 107(10):104038, 2023.
\newblock \href {http://arxiv.org/abs/2301.11480} {\path{arXiv:2301.11480}},
  \href {https://doi.org/10.1103/PhysRevD.107.104038}
  {\path{doi:10.1103/PhysRevD.107.104038}}.

\end{thebibliography}

\end{document}